\documentclass[aip,preprint,jcp,showpacs,amsmath,amssymb,superscriptaddress,a4paper]{revtex4-1} 

\usepackage{graphicx}
\usepackage{bm}

\begin{document}

\preprint{}

\title{Electronic ground states of Fe$_2^+$ and Co$_2^+$ as determined by x-ray absorption and x-ray magnetic circular dichroism spectroscopy}

\author{V.~Zamudio-Bayer}
 \affiliation{Institut f\"ur Methoden und Instrumentierung der Forschung mit Synchrotronstrahlung, Helmholtz-Zentrum Berlin f\"ur Materialien und Energie, Albert-Einstein-Stra{\ss}e 15, 12489 Berlin, Germany}
 \affiliation{Physikalisches Institut, Universit\"{a}t Freiburg, Stefan-Meier-Stra{\ss}e 21, 79104 Freiburg, Germany}

\author{K.~Hirsch}
\author{A.~Langenberg}
 \affiliation{Institut f\"ur Methoden und Instrumentierung der Forschung mit Synchrotronstrahlung, Helmholtz-Zentrum Berlin f\"ur Materialien und Energie, Albert-Einstein-Stra{\ss}e 15, 12489 Berlin, Germany}
 \affiliation{Institut f\"{u}r Optik und Atomare Physik, Technische Universit\"{a}t Berlin, Hardenbergstra{\ss}e 36, 10623 Berlin, Germany}

\author{A.~{\L}awicki}
 \affiliation{Institut f\"ur Methoden und Instrumentierung der Forschung mit Synchrotronstrahlung, Helmholtz-Zentrum Berlin f\"ur Materialien und Energie, Albert-Einstein-Stra{\ss}e 15, 12489 Berlin, Germany}

\author{A.~Terasaki}
 \affiliation{Cluster Research Laboratory, Toyota Technological Institute, 717-86 Futamata, Ichikawa, Chiba 272-0001, Japan}
 \affiliation{Department of Chemistry, Kyushu University, 744 Motooka, Nishi-ku, Fukuoka 819-0395, Japan}

\author{B.~v.~Issendorff}
 \affiliation{Physikalisches Institut, Universit\"{a}t Freiburg, Stefan-Meier-Stra{\ss}e 21, 79104 Freiburg, Germany}

\author{J.~T.~Lau}
 \email{tobias.lau@helmholtz-berlin.de}
 \affiliation{Institut f\"ur Methoden und Instrumentierung der Forschung mit Synchrotronstrahlung, Helmholtz-Zentrum Berlin f\"ur Materialien und Energie, Albert-Einstein-Stra{\ss}e 15, 12489 Berlin, Germany}

\date{\today}

\begin{abstract}
The $^6\Pi$ electronic ground state of the Co$_2^+$ diatomic molecular cation has been assigned experimentally by x-ray absorption and x-ray magnetic circular dichroism spectroscopy in a cryogenic ion trap. Three candidates, $^6\Phi$, $^8\Phi$, and $^8\Gamma$, for the electronic ground state of Fe$_2^+$ have been identified. 
These states carry sizable orbital angular momenta that disagree with theoretical predictions from multireference configuration interaction and density functional theory. 
Our results show that the ground states of neutral and cationic diatomic molecules of $3d$ transition elements cannot generally be assumed to be connected by a one-electron process. 
\end{abstract}


\maketitle

\section{Introduction}

Spin correlation and orbital angular momentum in the open-shell $3d$ transition metals lead to a large variety of electronic ground states, magnetic coupling, or chemical reactivity already in diatomic molecules.~\cite{Morse86,Barden00,Lombardi02,Gutsev03a} 
The complexity of the electronic structure of $3d$ diatomics is further reflected in a number of conflicting results from theoretical predictions and experimental studies \cite{Lin69,Purdum82,Rohlfing84,Baumann84,Leopold86,Leopold88,VanZee92,Danset04} on the electronic ground states of Fe$_2$ and Co$_2$ that have extensively been discussed in the literature.~\cite{Morse86,Barden00,Lombardi02,Gutsev03a}  
The current agreement among theory for neutral Fe$_2$ and Co$_2$ diatomic molecules is on $^9\Sigma_g^-$ and $^5\Delta_g$ ground states, respectively.~\cite{Huebner02,Irigoras03,Wang05e,Rollmann06,Strandberg07,Fritsch08,Blonski09,Xiao09,Angeli11,Hoyer14,Kalemos15} 
Experimental data, on the other hand, has resisted to yield a conclusive picture.~\cite{Lin69,Purdum82,Rohlfing84,Baumann84,Leopold86,Leopold88,VanZee92,Danset04}
From a different point of view, the possibility of aligning molecules in magnetic fields \cite{Friedrich92,Slenczka94} and the question of the fundamental limits of the magnetic anisotropy energy that is responsible for a preferred axis of alignment of the magnetic moment naturally lead to diatomic molecules of $3d$ transition elements and their complexes.~\cite{Strandberg07,Fritsch08,Blonski09,Xiao09} 
Again, the predicted magnetic anisotropy energy relies on an accurate determination of the electronic states and their relative energies. 
\newline
At present there is only limited experimental information on the spin and orbital magnetic moments of gaseous $3d$ transition-metal molecules or molecular ions. In particular for free Fe$_2^+$ and Co$_2^+$ only the bond dissociation energies are known experimentally~\cite{Loh89,Lian92a,Russon93,Russon94,Hales94}
Available theoretical results predict $^8\Sigma$, $^{10}\Sigma$, or $^8\Delta$ ground states \cite{Gutsev03a,Irigoras03,Rollmann06a,Chiodo06,Hoyer14,Kalemos15} for Fe$_2^+$ as well as $^6\Sigma$ or $^6\Gamma$ ground states \cite{Jamorski97,Gutsev03a} for Co$_2^+$.
\newline
In conflict with these predictions, we here present $^6\Phi$, $^8\Phi$, and $^8\Gamma$ candidates for the electronic ground state of Fe$_2^+$, as well as the $^6\Pi$ electronic ground state of Co$_2^+$ as determined by gas-phase x-ray magnetic circular dichroism (XMCD) spectroscopy~\cite{Peredkov11b,Niemeyer12,ZamudioBayer13,Langenberg14,Hirsch15a,ZamudioBayer15a,ZamudioBayer15b} that was only recently introduced for free molecular ions.~\cite{ZamudioBayer15a,ZamudioBayer15b} 
All states that are found in our study carry significant orbital angular momentum that disagrees with theoretical preferences for $\Sigma$ states.

\section{Experimental Setup and Methods}

Molecular ions are produced by magnetron sputtering of high-purity (99.95~\%) iron or cobalt sputtering targets with argon (99.9999~\%) at $\approx 100$~sccm flow rate and helium (99.9999~\%) buffer gas at $\approx 500$~sccm flow rate in a liquid-nitrogen cooled gas-aggregation cluster source.~\cite{Hirsch09} During formation, these ions are cooled to $\approx 150$ K by multiple collisions with buffer gas atoms at $p\approx 1$~mbar stagnation pressure. This allows for depopulation of excited states. 
The ions are collected at the source exit by a radio-frequency hexapole ion guide and are transmitted through a differential pumping stage into a radio-frequency quadrupole mass filter. After mass selection, Fe$_2^+$ or Co$_2^+$ diatomic molecular cations are accumulated in a liquid-helium cooled linear ion trap and are thermalized to $T = 10 - 20$~K by collisions with helium buffer gas at $p \approx 10^{-3}$~mbar, equivalent to a helium atom number density of $n \approx 10^{14}$~cm$^{-3}$. 
The ion trap housing was kept at a temperature of $\approx 4$~K but the resulting ion temperature will in any case be higher than the ion trap housing or electrode temperature because of inevitable radio-frequency heating \cite{Major68} that competes with collisional cooling and sensitively depends on buffer gas and radio-frequency parameters. 
Further relaxation of any remaining excited electronic state is achieved by typical storage (half-life) times of the parent ions on the order of $1 - 10$~s in the ion trap.~\cite{Hirsch12a,ZamudioBayer15a,ZamudioBayer15b} 
\newline
An elliptically polarized and monochromatic soft x-ray beam, delivered by variable polarization undulator beam line UE52-PGM/SGM of the synchrotron radiation facility BESSY~II, is coupled in along the axis of the ion trap. 
The incident photon energy was scanned across the $L_{2,3}$ absorption edges to probe $2p \rightarrow 3d$ transitions of iron or cobalt with a photon energy bandwidth of 625~meV in 250~meV steps. 
$L_{2,3}\, (2p \rightarrow 3d)$ core excitation is followed by Auger decay cascades that lead to multiple ionization and dissociation of the excited parent ion.
The most intense product ion that results from $2p$ core excitation and relaxation is the atomic dication for both Fe$_2^+$ and Co$_2^+$ parent ions at our experimental conditions where collisions with helium atoms can quench higher charge states $(q \ge 3)$ of Fe$^{q+}$ and Co$^{q+}$ product ions by Penning ionization. 
This Fe$^{2+}$ or Co$^{2+}$ ion yield is recorded at every photon energy step with an in-line reflectron time-of-flight mass spectrometer and is normalized to the incident photon flux as detected with a GaAsP photodiode. 
\newline
For XMCD spectroscopy, the ion trap is placed inside the homogeneous magnetic field ($\mu_0H = 5$~T) of a superconducting solenoid.~\cite{Terasaki07,Niemeyer12,ZamudioBayer13,Langenberg14,Hirsch15a,ZamudioBayer15a,ZamudioBayer15b} 
Following the standard procedures, the XMCD spectrum is obtained as the difference of x-ray absorption spectra that are recorded with parallel and antiparallel orientation of the photon helicity relative to the direction of the applied magnetic field. The isotropic x-ray absorption spectrum is obtained by taking the average of the spectra of both photon helicities. 
XMCD sum rules \cite{Thole92,Carra93} are applied to these spectra to derive expectation values of spin and orbital angular momentum in the ensemble average. 
\newline
While the orbital angular momentum sum rule \cite{Thole92} of x-ray magnetic circular dichroism can be applied without complications, the spin sum rule \cite{Carra93} potentially contains contributions of the magnetic dipole term $T_z$ and therefore delivers an effective spin magnetic moment $m_s^{\text{eff}} = m_s + m_T$ with 
$\langle m_T \rangle = (7 / 2) (g \mu_B / \hbar)\, \langle T_z \rangle \approx 7 (\mu_B / \hbar)\, \langle T_z \rangle$. 
This magnetic dipole term $T_z$ is a second order multipole contribution that is related to the difference of the quadrupole moments of the charge distributions for spin-up and spin-down electrons. 
A sizable $T_z$ contribution to the spin sum rule can be observed if the absolute value of $\langle T_z \rangle$ is large and the experiment is performed in a non-isotropic geometry. This is typically the case for surfaces and thin films.~~\cite{Wu94,Stoehr95a,Stoehr95b,Duerr96a,vanderLaan14}

\section{X-ray absorption and XMCD spectra of {Fe}$\bm{_2^+}$ and {Co}$\bm{_2^+}$}

\begin{figure}[t]
	\includegraphics{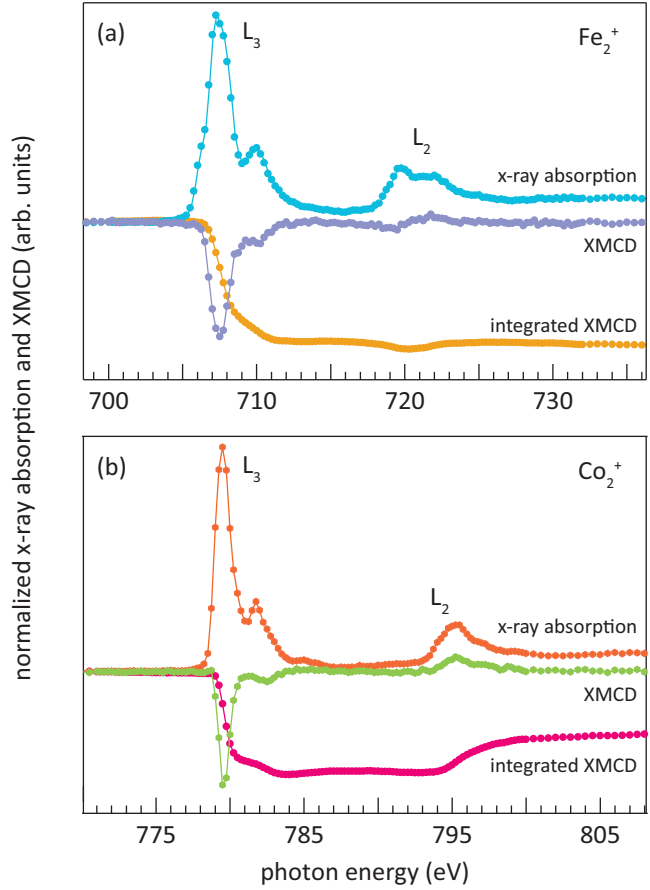}
	\caption{\label{fig:XS_Fe2andCo2} $L_{2,3}$ X-ray absorption, x-ray magnetic circular dichroism (XMCD), and integrated XMCD spectra of (a) Fe$_2^+$ and (b) Co$_2^+$ at $\mu_0H = 5$ T in a 4 K ion trap. The x-ray absorption and XMCD spectra as the respective average and difference of the spectra for positive and negative photon helicity are displayed on the same scale within each panel, normalized to the non-resonant photoionization intensity at photon energies well above the $L_2$-edge. 
These raw XMCD spectra reflect the magnetization per unoccupied $3d$ state at our experimental conditions but not the total magnetic moment. The large value of the integrated XMCD above the $L_2$ absorption edge indicates significant orbital magnetization in both cases.}
\end{figure}

In Fig.\ \ref{fig:XS_Fe2andCo2} the $L_{2,3}$ x-ray absorption and XMCD spectra of Fe$_2^+$ and Co$_2^+$ at $B = 5$~T in the 4~K ion trap are shown along with the XMCD spectra, integrated with respect to photon energy, to which the XMCD sum rules \cite{Thole92,Carra93} are applied.
The spectral shape of x-ray absorption in both cases is identical to that recorded at higher ion trap temperature of $\approx 90$~K without applied magnetic field,~\cite{Hirsch09,Hirsch12b} i.e., no temperature or magnetic field dependence of the x-ray absorption spectrum was observed. 
Even though some multiplet structure \cite{deGroot05} is still visible in the $L_{2,3}$ x-ray absorption spectra of the diatomic molecular cations, it is not as well resolved as in the corresponding atomic or cationic spectra.~\cite{Martins03,Richter04a,Hirsch12a} This indicates delocalization of $3d$ orbitals and the formation of molecular orbitals with $3d$ participation.
\newline
Application of the XMCD sum rules~\cite{Thole92,Carra93} to these spectra results in orbital $\left\langle m_L\right\rangle$ and effective spin $\left\langle m_S^{\text{eff}}\right\rangle$ magnetizations per unoccupied $3d$ state at our experimental conditions of 
$\left\langle m_L \right\rangle = (0.24 \pm 0.01)$~$\mu_\mathrm{B}$ 
and 
$\left\langle m_S^{\text{eff}}\right\rangle = (0.42 \pm 0.02)$~$\mu_\mathrm{B}$ 
for Fe$_2^+$, and  
$\left\langle m_L \right\rangle = (0.11 \pm 0.01)$~$\mu_\mathrm{B}$ 
and 
$\left\langle m_S^{\text{eff}}\right\rangle = (0.47 \pm 0.08)$~$\mu_\mathrm{B}$ 
for Co$_2^+$. 
Already at this stage our experimental study shows significant orbital contributions to the total magnetization and clearly excludes $\Sigma$ ground states for Fe$_2^+$ and Co$_2^+$.
\newline
Inherent to the application of XMCD sum rules, the experimental ratios of orbital-to-spin magnetic moments are only limited by statistical uncertainties and a possible contribution of the magnetic dipole term, but not by the number $n_h$ of unoccupied $3d$ states, the ion temperature, or the degree of circular polarization of the incident photon beam. Because of the high confidence in this value, the following analysis of the electronic ground states will be primarily based on this ratio of orbital-to-spin magnetization. Our experimental data yields 
values of 
$\left\langle m_L\right\rangle / \left\langle m_S^{\text{eff}}\right\rangle = 0.57\pm0.04$ 
for Fe$_2^+$ and 
$\left\langle m_L\right\rangle / \left\langle m_S^{\text{eff}}\right\rangle = 0.24\pm0.04$ 
for Co$_2^+$.

\section{Electronic ground states of Fe$\bm{_2^+}$ and Co$\bm{_2^+}$}

\subsection{Electronic configurations and number of unoccupied $\bm{3d}$ states}

In general, a $(4s\,\sigma)^2\,(3d\,\sigma,\pi,\delta)^{13}$ and a $(4s\,\sigma)^2\,(3d\,\sigma,\pi,\delta)^{15}$ molecular configuration can be assumed for Fe$_2^+$ and Co$_2^+$, respectively, since a molecular $(4s\,\sigma)^2\,(3d\,\sigma,\pi,\delta)^{n}$ configuration with a two-electron $\sigma$ bond should be energetically favorable over a $(4s\,\sigma)^3\,(3d\,\sigma,\pi,\delta)^{n-1}$ configuration with an additional singly occupied antibonding $4s\,\sigma^*$ orbital.~\cite{Rohlfing84,Loh88,Loh89,Russon93,Russon94,Conceicao96,Strandberg07} 
These molecular electronic configurations correspond to $n_h(\mathrm{Fe}_2^+) = 3.5$ and $n_h(\mathrm{Co}_2^+) = 2.5$ unoccupied $3d$ states per atom. 
With these values and the experimentally determined magnetization per $3d$ hole, we obtain the magnetization per atom as 
$\left\langle m_L \right\rangle = (0.83 \pm 0.02)$~$\mu_\mathrm{B}$ 
and 
$\left\langle m_S^{\text{eff}}\right\rangle = (1.47 \pm 0.08)$~$\mu_\mathrm{B}$ 
for Fe$_2^+$, and  
$\left\langle m_L \right\rangle = (0.29 \pm 0.02)$~$\mu_\mathrm{B}$ 
and 
$\left\langle m_S^{\text{eff}}\right\rangle = (1.18 \pm 0.21)$~$\mu_\mathrm{B}$ 
for Co$_2^+$. 
The detected molecular orbital magnetization of $(1.67 \pm 0.03)$~$\mu_\mathrm{B}$ immediately rules out $\Sigma$ and $\Pi$ states for Fe$_2^+$. Likewise, $\Sigma$ states are incompatible with the experimental molecular orbital magnetization of $(0.57 \pm 0.05)$~$\mu_\mathrm{B}$ for Co$_2^+$.

\subsection{Identification of candidate states and assignment of the electronic ground states of Fe$\bm{_2^+}$ and Co$\bm{_2^+}$}

\subsubsection{Identification of candidate states compatible with the experimental ratio of orbital-to-spin magnetization}

\begin{figure}[t]
	\includegraphics{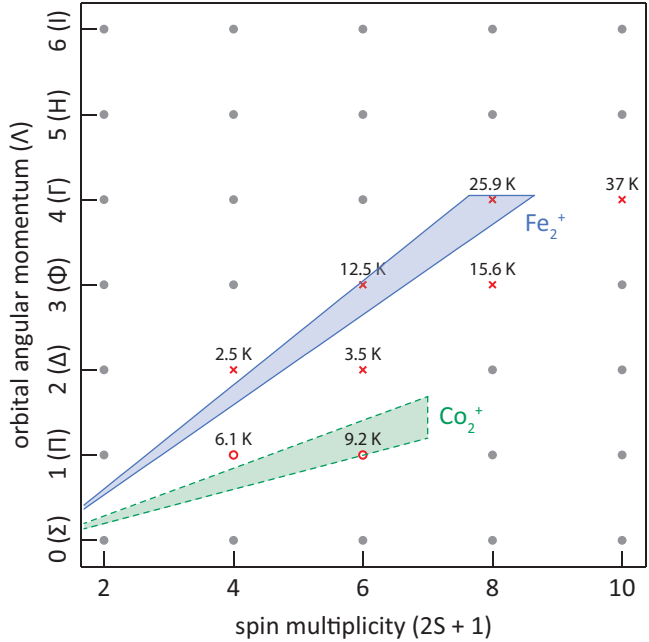}
	\caption{\label{fig:LtoSratios} Spin multiplicities $2 S + 1$  and orbital angular momenta $\Lambda$ (filled grey circles) that are allowed for the Fe$_2^+$ and Co$_2^+$. The spin multiplicity is limited to $2 S + 1 \le 6$ for Co$_2^+$, and the orbital angular momentum to $\Lambda \le 4$ for Fe$_2^+$.
Solid blue and dashed green lines (shaded areas) confine the combinations of $2 S + 1$ and $\Lambda$ that are compatible within one standard deviation with the experimental $m_L / m_S^{\text{eff}}$ ratios given by the XMCD sum rules for Fe$_2^+$ (red crosses) and Co$_2^+$ (open red circles). 
Symbols outside the error bars mark states that are taken into account because of a possible magnetic dipole $(T_z)$ contribution to $m_S^{\text{eff}}$. Numerical values are the ion temperatures that would correspond to the candidate states.}
\end{figure}

Following the Wigner-Witmer correlation rules, \cite{Wigner28} only a limited number of combinations of molecular spin ($S$) and orbital ($\Lambda$) quantum numbers are compatible with the atomic and ionic ground state dissociation asymptotes and with the experimentally determined values for the orbital-to-spin magnetic moment ratio in the given molecular electronic configurations.  
Allowed values of the spin multiplicity  and orbital angular momentum are 
$2 S + 1 \in \{2,4,6,8,10\}$, 
$\Lambda \in \{0 (\Sigma),1 (\Pi),2 (\Delta),3 (\Phi),4 (\Gamma)\}$ 
for Fe$_2^+$ with the Fe $(^5D_4)$ $+$ Fe$^+$ $(^6D_{9/2})$ asymptote,  and 
$2 S + 1 \in \{2,4,6\}$, 
$\Lambda \in \{0 (\Sigma),1 (\Pi),2 (\Delta),3 (\Phi),4 (\Gamma),5 (\text{H}), 6 (\text{I})\}$ 
for Co$_2^+$ with the Co $(^4F_{9/2})$ $+$ Co$^+$ $(^3F_4)$ asymptote. 
These states are considered in Fig.\ \ref{fig:LtoSratios}, where the orbital angular momentum $\Lambda$ is plotted versus the spin multiplicity $2 S + 1$ for comparison with the experimental ratios of orbital-to-spin magnetization. 
As can be seen from this figure there are two states, $^6\Phi$ and $^8\Gamma$, in the case of Fe$_2^+$ that are within the range of the experimental values whereas for Co$_2^+$ only the $^6\Pi$ state falls directly within the range of the experimental data. However, more states have to be considered if a non-vanishing contribution of the magnetic dipole term $T_z$ to the effective spin is to be taken into account.

\subsubsection{Contribution of the magnetic dipole term $\bm{T_z}$ to the effective spin magnetic moment}

A strong magnetic field at low temperature leads to an alignment of diatomic molecular ions with nonvanishing orbital angular momentum.~\cite{Friedrich92,Slenczka94} 
In this anisotropic situation, possible contributions of the magnetic dipole term $T_z$ to the effective spin magnetization \cite{Carra93} have to be considered, because these could affect the experimental ratio of orbital-to-effective-spin magnetization.~\cite{Wu94,Stoehr95a,Stoehr95b,Duerr96a,vanderLaan14}
In cylindrical symmetry, the magnetic dipole term has components parallel ($T_\parallel$) and perpendicular ($T_\perp$) to the molecular axis, with $2\, T_\perp = - T_\parallel$. These will cancel out in the angular average of freely rotating states. In non-fully aligned pendular states \cite{Friedrich92,Slenczka94}, both $T_\parallel$ and $T_\perp$ will contribute to the effective spin with their opposite signs, resulting in a low effective $T_z$.  
\newline
To estimate a possible contribution of $T_z$ to $m_S^{\text{eff}}$, we start from the maximum value of $T_z$ that is present in the free atom. 
This value will be reduced by the formation of bonds. We therefore estimate $T_z$ in the molecular ion to be $\le 2/3$ of the average of the absolute values in the asymptotic atomic and ionic contributions, which are
$7\, \langle T_z \rangle = - 2 \hbar$ for $3d^6$; 
$7\, \langle T_z \rangle = - \hbar$ for $3d^7$; 
and 
$7\, \langle T_z \rangle = \hbar$ for $3d^8$ 
configurations.~\cite{Crocombette96a,Piamonteze09}
In addition, we take an incomplete alignment of the molecular axis into account by further reduction to $\le 1/2$ of these values.
We thus consider 
$- \hbar / 2 \le 7\,  T_z \le  \hbar / 2$ 
per atom for Fe$_2^+$ and 
$- \hbar / 3 \le 7\,  T_z \le \hbar / 3$ 
per atom for Co$_2^+$. These estimates are in agreement with predictions for $T_z$ in free-standing and supported monatomic wires.~\cite{Komelj02,Ederer03,Ederer03a,Minar06,Sipr09a} 
\newline
An additional complication could arise from non-integer occupation numbers in the $3d$ derived states because of a possible mixing of $3d\, \sigma$ and $4s\, \sigma$ orbitals. 
Since only spin but no orbital angular momentum can be transferred between $3d\, \sigma$ and $4s\, \sigma$ states, this would again affect the observed ratio of orbital-to-effective-spin magnetization. Electron transfer between $3d\, \sigma$ and $4s\, \sigma$ states could occur in either direction and can be taken into account as an additional variation $S_{3d-4s}$ of the effective spin. We estimate the transfer of 0.5 electrons as an upper limit per molecule, equivalent to an additional contribution of $- \hbar / 4 \le S_{3d-4s} \le \hbar / 4$ to the effective spin per atom, which is added to the $7\,  T_z$ contribution. 
This leads to a total of 
$- 3 \hbar / 4 \le 7\,  T_z + S_{3d-4s} \le  3 \hbar / 4$ 
per atom for Fe$_2^+$ and 
$- 7 \hbar / 12 \le 7\, T_z + S_{3d-4s} \le 7 \hbar / 12$ 
per atom for Co$_2^+$. 
\newline
With this estimate, the 
$^4\Delta$ ($7\,  T_z + S_{3d-4s} = 0.14~\hbar$),
$^6\Delta$ ($7\,  T_z + S_{3d-4s} = - 0.61~\hbar$),
$^8\Phi$ ($7\,  T_z + S_{3d-4s} = -0.67~\hbar$), and
$^{10}\Gamma$ ($7\,  T_z + S_{3d-4s} = - 0.73~\hbar$),
states of Fe$_2^+$ as well as the 
$^4\Pi$ ($7\,  T_z + S_{3d-4s} = 0.29~\hbar$) 
state of Co$_2^+$
would have to be considered as additional candidates for the electronic ground states. 
\newline
Our estimate is a conservative one because it takes into account both, maximum $T_z$ and maximum $3d\, \sigma / 4s\, \sigma$ electron transfer contributions at the same time. 
For Co$_2^+$, one might even expect a cancellation of $T_z$ because of the opposite signs of $T_z$ in $3d^7$ and $3d^8$ configurations.~\cite{Crocombette96a,Piamonteze09}

\subsubsection{Assessment of candidate states}

All candidate states can be assessed by the ion temperature at which they would match the experimentally determined magnetization. 
This ion temperature  is obtained from an effective Zeeman-Hamiltonian \cite{Berdyugina02,AsensioRamos06} in Hund's coupling case (a) of the rotating molecular ion in a magnetic field. 
An equilibrium distance of 2.0~{\AA} was used to estimate the rotational constants. This estimate is based on the existing experimental values \cite{Asher94a,Yang00b} of $r_e(\text{V}_2^+) = 1.73$~{\AA} and $r_e(\text{Ni}_2^+) = 2.22$~{\AA} and predicted values \cite{Jamorski97,Gutsev03a,Irigoras03,Rollmann06a,Chiodo06,Hoyer14,Kalemos15} of $r_e(\text{Fe}_2^+) = 2.1 - 2.2$ {\AA} and $r_e(\text{Co}_2^+) = 1.9 - 2.1$ {\AA}, for which no experimental data exist. The $^{10}\Sigma$ state of Fe$_2^+$ that is predicted to have an equilibrium distance of $2.7 - 3.0$~{\AA}, significantly larger than in the other states,~\cite{Irigoras03, Chiodo06} is not considered here since we can clearly rule out $\Sigma$ states from our non-vanishing orbital magnetization. 
Hund's case (a) should apply here \cite{LefebvreBrion86} because of non-vanishing orbital angular momentum, a spin-orbit coupling constant $\zeta$ on the order of 10 meV, and a rotational constant $B$ on the order of 10 $\mu$eV such that $\zeta \Lambda \gg B J$ at our experimental conditions.
Because of the smaller experimental error for orbital than for spin magnetization and because of a possible $T_z$ contribution to the effective spin sum rule, we have calculated $\langle m_J \rangle$ as 
$\langle m_J \rangle = \langle m_L \rangle \left(1 + 2 S / \Lambda \right)$ 
for each $^{2 S + 1}\Lambda$ candidate state rather than 
$\langle m_J \rangle = \langle m_L \rangle + \langle m_S^{\text{eff}} \rangle$ 
to determine the ion temperature in the following. 
\newline
For Fe$_2^+$ the ion temperatures that would correspond to the candidate states are
$T(^4\Delta) = (2.5 \pm 0.7)$~K, 
$T(^6\Delta) = (3.5 \pm 0.9)$~K, 
$T(^6\Phi) = (12.5 \pm 0.9)$~K, 
$T(^8\Phi) = (15.6 \pm 1.0) $~K, 
$T(^8\Gamma) = (25.9 \pm 1.4) $~K, and 
$T(^{10}\Gamma) = (37 \pm 2) $~K. 
For Co$_2^+$ the corresponding ion temperatures are 
$T(^4\Pi) = (6.1 \pm 0.8)$~K, and	
$T(^6\Pi) = (9.2 \pm 1.2)$~K. 
The $^4\Delta$ and $^6\Delta$ states of Fe$_2^+$ can be discarded as they would correspond to ion temperatures below the ion trap temperature of 4~K. Likewise, the $^4\Pi$ state of Co$_2^+$  would correspond to an unrealistically low radio-frequency heating of only 2~K, whereas the $^{10}\Gamma$ state of Fe$_2^+$ would indicate unrealistically high radio-frequency heating of 33~K. 
Of the remaining states, $^6\Phi$ and $^8\Phi$ for Fe$_2^+$ and $^6\Pi$ for Co$_2^+$ correspond to a radio-frequency heating of $5 - 12$~K, which is in reasonable agreement with what we have observed for the Cr$_2^+$, Mn$_2^+$, and Mn$_3^+$ molecular ions at similar experimental conditions.~\cite{ZamudioBayer15a, ZamudioBayer15b} 
The $^8\Gamma$ state of Fe$_2^+$ would correspond to a radio-frequency heating of 22~K, which is higher than expected at our experimental conditions but cannot be ruled out. 
Interestingly, the $^6\Phi$ state of Fe$_2^+$ with $n_h = 3.5$ unoccupied $3d$ derived states per atom can only be reached if there is an unoccupied majority spin state, i.e., if the spin polarization per $3d$ hole is $5/7 < 1$. For Co$_2^+$ with $n_h = 2.5$ the majority spin states are completely filled and the spin polarization per unoccupied $3d$ derived state is equal to 1, as would also be the case for the $2 S + 1 = 8$ spin states of Fe$_2^+$. 
\newline
In summary, we have determined the ground state of Co$_2^+$ as $^6\Pi$. For Fe$_2^+$ we can narrow down the range of candidates for the ground state to $^6\Phi$, $^8\Phi$, and $^8\Gamma$.

\section{Comparison to theoretical predictions and previous experimental results}

Even though diatomic molecules and molecular ions of the $3d$ transition elements have been extensively treated computationally over the last decades,~\cite{Morse86,Barden00,Lombardi02,Gutsev03a} often as benchmark systems, the ground states of many of these species have not been identified unambiguously because of computational challenges and because of the lack of experimental confirmation. 
Most of the available theoretical studies on iron and cobalt diatomics treat neutral molecules, with fewer predictions available for the diatomic cations. One assumption that is often made to lower the computational demand when predicting the electronic ground state of diatomic molecular cations is that the spin multiplicities of the neutral and the cationic ground state differ only by $\pm 1$, i.e. that the ground state of the molecular cation can be reached from the neutral molecule in a one-electron process.

\subsection{Predicted ground states of Fe$\bm{_2^+}$}

The di-iron molecular cation has been modeled by density functional theory at different levels, which predict either an $^8\Delta$ ground state \cite{Gutsev03a,Rollmann06a} or a $^{10}\Sigma$ ground state \cite{Irigoras03,Chiodo06}, whereas recent multireference configuration interaction studies of Fe$_2^+$ predict an $^8\Sigma$ ground state.~\cite{Hoyer14,Kalemos15} 
Comparison with our experimentally determined $^6\Phi$, $^8\Phi$, and $^8\Gamma$ candidate states shows that theory underestimates the orbital degeneracy of the ground state in the case of Fe$_2^+$. 
\newline
For density functional theory this discrepancy might be attributed to the difficulty of treating orbital angular momentum and electron correlation in $3d$ transition elements, while  in the multireference configuration interaction approach the reason might be in the restriction to $2 S + 1 = 8$ and $2 S + 1 = 10$ spin states \cite{Hoyer14,Kalemos15} when taking the $2 S + 1 = 9$ spin state of neutral Fe$_2$ and the assumption of ionization as a one-electron process a starting points. 
However, none of our experimentally determined candidate states can be reached from the currently accepted $^9\Sigma_g^-$ ground state \cite{Huebner02,Angeli11,Hoyer14,Kalemos15} of neutral Fe$_2$ in a one-electron transition. Similar results have also been obtained experimentally for the cases of chromium and  manganese.~\cite{Niemeyer12,Langenberg14,ZamudioBayer15a,ZamudioBayer15b,Egashira15} 
We therefore propose to follow an unrestricted and unbiased theoretical approach in the search of the electronic ground state.

\subsection{Predicted ground states of Co$\bm{_2^+}$}

In contrast to Fe$_2^+$ for which the only experimental data that is available is the bond dissociation energy \cite{Loh88,Lian92a}, an attempt has been made to assign the electronic ground state of Co$_2^+$ from matrix-isolation electron spin resonance spectroscopy at 2~K in a neon matrix.~\cite{VanZee92} 
The observed electron spin resonance signal was interpreted to arise from a $^6\Sigma$ state which agrees on the multiplicity but disagrees on the orbital degeneracy with the $^6\Pi$ state identified as the ground state in our work. This discrepancy might arise from the selective sensitivity of electron spin resonance spectroscopy to an excited $\Sigma$ state of Co$_2^+$ as this state was only observed if the matrix was not annealed.~\cite{VanZee92}
\newline
The few existing density functional theory studies on cationic Co$_2^+$ consistently predict a ground-state spin multiplicity of $2 S + 1 = 6$ in agreement with the experimental results presented here. However, the calculations predict $\Sigma$ or $\Gamma$ symmetry of the ground state.~\cite{Jamorski97,Gutsev03a} 
Both symmetries can be excluded from our experimental result.
\newline
In contrast to the case of Fe$_2^+$, the $^6\Pi$ ground state of Co$_2^+$ could in principle be connected to the predicted $^5\Delta$ ground state \cite{Wang05e,Strandberg07,Fritsch08,Blonski09} of neutral Co$_2$ via a one-electron process, namely by the removal of a $3d\, \pi$ minority electron.

\section{Conclusion and outlook}

The experimentally determined $^6\Pi$ ground state of Co$_2^+$ as well as the $^6\Phi$, $^8\Phi$, and $^8\Gamma$ candidates for the ground state of Fe$_2^+$ disagree with state-of-the art multireference configuration interaction \cite{Hoyer14,Kalemos15} and density functional theory \cite{Jamorski97,Gutsev03a,Irigoras03,Chiodo06,Rollmann06a} predictions. 
These results underline the importance of electronic correlation effects and orbital angular momentum in $3d$ transition elements.
Our results add further evidence to the finding that the ground states of neutral and cationic diatomic molecules of the $3d$ transition elements are not necessarily connected by one-electron processes. Instead, spin or orbital angular momentum can change significantly upon ionization. 
This is similar to the configuration change in the atomic and cationic ground states of the $3d$ elements vanadium, cobalt, and nickel that are not linked by one-electron processes either.  
The restriction to states that can be directly connected to the ground state should therefore be lifted in theoretical approaches for an unbiased search of the ground state even though this will increase the computational demand.

\begin{acknowledgments}
Beam time for this project was granted at BESSY~II beamlines UE52-SGM and UE52-PGM, operated by Helmholtz-Zentrum Berlin. Skillful technical assistance and support by Thomas Blume, Robert Schulz, Helmut Pfau, and Fran\c{c}ois Talon are gratefully acknowledged. This project was partially funded by the German Federal Ministry of Education and Research (BMBF) through grant BMBF-05K13Vf2. The superconducting solenoid was kindly provided by Toyota Technological Institute. AT acknowledges financial support by Genesis Research Institute, Inc. BvI acknowledges travel support by Helmholtz-Zentrum Berlin. 
\end{acknowledgments}


\begin{thebibliography}{71}%
\makeatletter
\providecommand \@ifxundefined [1]{%
 \@ifx{#1\undefined}
}%
\providecommand \@ifnum [1]{%
 \ifnum #1\expandafter \@firstoftwo
 \else \expandafter \@secondoftwo
 \fi
}%
\providecommand \@ifx [1]{%
 \ifx #1\expandafter \@firstoftwo
 \else \expandafter \@secondoftwo
 \fi
}%
\providecommand \natexlab [1]{#1}%
\providecommand \enquote  [1]{``#1''}%
\providecommand \bibnamefont  [1]{#1}%
\providecommand \bibfnamefont [1]{#1}%
\providecommand \citenamefont [1]{#1}%
\providecommand \href@noop [0]{\@secondoftwo}%
\providecommand \href [0]{\begingroup \@sanitize@url \@href}%
\providecommand \@href[1]{\@@startlink{#1}\@@href}%
\providecommand \@@href[1]{\endgroup#1\@@endlink}%
\providecommand \@sanitize@url [0]{\catcode `\\12\catcode `\$12\catcode
  `\&12\catcode `\#12\catcode `\^12\catcode `\_12\catcode `\%12\relax}%
\providecommand \@@startlink[1]{}%
\providecommand \@@endlink[0]{}%
\providecommand \url  [0]{\begingroup\@sanitize@url \@url }%
\providecommand \@url [1]{\endgroup\@href {#1}{\urlprefix }}%
\providecommand \urlprefix  [0]{URL }%
\providecommand \Eprint [0]{\href }%
\providecommand \doibase [0]{http://dx.doi.org/}%
\providecommand \selectlanguage [0]{\@gobble}%
\providecommand \bibinfo  [0]{\@secondoftwo}%
\providecommand \bibfield  [0]{\@secondoftwo}%
\providecommand \translation [1]{[#1]}%
\providecommand \BibitemOpen [0]{}%
\providecommand \bibitemStop [0]{}%
\providecommand \bibitemNoStop [0]{.\EOS\space}%
\providecommand \EOS [0]{\spacefactor3000\relax}%
\providecommand \BibitemShut  [1]{\csname bibitem#1\endcsname}%
\let\auto@bib@innerbib\@empty
\bibitem [{\citenamefont {Morse}(1986)}]{Morse86}%
  \BibitemOpen
  \bibfield  {author} {\bibinfo {author} {\bibfnamefont {M.~D.}\ \bibnamefont
  {Morse}},\ }\href {\doibase 10.1021/cr00076a005} {\bibfield  {journal}
  {\bibinfo  {journal} {Chem. Rev.}\ }\textbf {\bibinfo {volume} {86}},\
  \bibinfo {pages} {1049} (\bibinfo {year} {1986})}\BibitemShut {NoStop}%
\bibitem [{\citenamefont {Barden}, \citenamefont {Rienstra-Kiracofe},\ and\
  \citenamefont {{Schaefer III}}(2000)}]{Barden00}%
  \BibitemOpen
  \bibfield  {author} {\bibinfo {author} {\bibfnamefont {C.~J.}\ \bibnamefont
  {Barden}}, \bibinfo {author} {\bibfnamefont {J.~C.}\ \bibnamefont
  {Rienstra-Kiracofe}}, \ and\ \bibinfo {author} {\bibfnamefont {H.~F.}\
  \bibnamefont {{Schaefer III}}},\ }\href {\doibase 10.1063/1.481916}
  {\bibfield  {journal} {\bibinfo  {journal} {J. Chem. Phys.}\ }\textbf
  {\bibinfo {volume} {113}},\ \bibinfo {pages} {690} (\bibinfo {year}
  {2000})}\BibitemShut {NoStop}%
\bibitem [{\citenamefont {Lombardi}\ and\ \citenamefont
  {Davis}(2002)}]{Lombardi02}%
  \BibitemOpen
  \bibfield  {author} {\bibinfo {author} {\bibfnamefont {J.~R.}\ \bibnamefont
  {Lombardi}}\ and\ \bibinfo {author} {\bibfnamefont {B.}~\bibnamefont
  {Davis}},\ }\href {\doibase 10.1021/cr010425j} {\bibfield  {journal}
  {\bibinfo  {journal} {Chem. Rev.}\ }\textbf {\bibinfo {volume} {102}},\
  \bibinfo {pages} {2431} (\bibinfo {year} {2002})}\BibitemShut {NoStop}%
\bibitem [{\citenamefont {Gutsev}\ and\ \citenamefont {{Bauschlicher,
  Jr.}}(2003)}]{Gutsev03a}%
  \BibitemOpen
  \bibfield  {author} {\bibinfo {author} {\bibfnamefont {G.~L.}\ \bibnamefont
  {Gutsev}}\ and\ \bibinfo {author} {\bibfnamefont {C.~W.}\ \bibnamefont
  {{Bauschlicher, Jr.}}},\ }\href {\doibase 10.1021/jp030146v} {\bibfield
  {journal} {\bibinfo  {journal} {J. Phys. Chem. A}\ }\textbf {\bibinfo
  {volume} {107}},\ \bibinfo {pages} {4755} (\bibinfo {year}
  {2003})}\BibitemShut {NoStop}%
\bibitem [{\citenamefont {Lin}\ and\ \citenamefont {Kant}(1969)}]{Lin69}%
  \BibitemOpen
  \bibfield  {author} {\bibinfo {author} {\bibfnamefont {S.-S.}\ \bibnamefont
  {Lin}}\ and\ \bibinfo {author} {\bibfnamefont {A.}~\bibnamefont {Kant}},\
  }\href {\doibase 10.1021/j100727a068} {\bibfield  {journal} {\bibinfo
  {journal} {J. Phys. Chem.}\ }\textbf {\bibinfo {volume} {73}},\ \bibinfo
  {pages} {2450} (\bibinfo {year} {1969})}\BibitemShut {NoStop}%
\bibitem [{\citenamefont {Purdum}\ \emph {et~al.}(1982)\citenamefont {Purdum},
  \citenamefont {Montano}, \citenamefont {Shenoy},\ and\ \citenamefont
  {Morrison}}]{Purdum82}%
  \BibitemOpen
  \bibfield  {author} {\bibinfo {author} {\bibfnamefont {H.}~\bibnamefont
  {Purdum}}, \bibinfo {author} {\bibfnamefont {P.~A.}\ \bibnamefont {Montano}},
  \bibinfo {author} {\bibfnamefont {G.~K.}\ \bibnamefont {Shenoy}}, \ and\
  \bibinfo {author} {\bibfnamefont {T.}~\bibnamefont {Morrison}},\ }\href
  {\doibase 10.1103/PhysRevB.25.4412} {\bibfield  {journal} {\bibinfo
  {journal} {Phys. Rev. B}\ }\textbf {\bibinfo {volume} {25}},\ \bibinfo
  {pages} {4412} (\bibinfo {year} {1982})}\BibitemShut {NoStop}%
\bibitem [{\citenamefont {Rohlfing}\ \emph {et~al.}(1984)\citenamefont
  {Rohlfing}, \citenamefont {Cox}, \citenamefont {Kaldor},\ and\ \citenamefont
  {Johnson}}]{Rohlfing84}%
  \BibitemOpen
  \bibfield  {author} {\bibinfo {author} {\bibfnamefont {E.~A.}\ \bibnamefont
  {Rohlfing}}, \bibinfo {author} {\bibfnamefont {D.~M.}\ \bibnamefont {Cox}},
  \bibinfo {author} {\bibfnamefont {A.}~\bibnamefont {Kaldor}}, \ and\ \bibinfo
  {author} {\bibfnamefont {K.~H.}\ \bibnamefont {Johnson}},\ }\href {\doibase
  http://dx.doi.org/10.1063/1.448168} {\bibfield  {journal} {\bibinfo
  {journal} {J. Chem. Phys.}\ }\textbf {\bibinfo {volume} {81}},\ \bibinfo
  {pages} {3846} (\bibinfo {year} {1984})}\BibitemShut {NoStop}%
\bibitem [{\citenamefont {Baumann}, \citenamefont {Van~Zee},\ and\
  \citenamefont {Weltner}(1984)}]{Baumann84}%
  \BibitemOpen
  \bibfield  {author} {\bibinfo {author} {\bibfnamefont {C.~A.}\ \bibnamefont
  {Baumann}}, \bibinfo {author} {\bibfnamefont {R.~J.}\ \bibnamefont
  {Van~Zee}}, \ and\ \bibinfo {author} {\bibfnamefont {W.}~\bibnamefont
  {Weltner}},\ }\href {\doibase 10.1021/j150653a029} {\bibfield  {journal}
  {\bibinfo  {journal} {J. Phys. Chem.}\ }\textbf {\bibinfo {volume} {88}},\
  \bibinfo {pages} {1815} (\bibinfo {year} {1984})}\BibitemShut {NoStop}%
\bibitem [{\citenamefont {Leopold}\ and\ \citenamefont
  {Lineberger}(1986)}]{Leopold86}%
  \BibitemOpen
  \bibfield  {author} {\bibinfo {author} {\bibfnamefont {D.~G.}\ \bibnamefont
  {Leopold}}\ and\ \bibinfo {author} {\bibfnamefont {W.~C.}\ \bibnamefont
  {Lineberger}},\ }\href {\doibase http://dx.doi.org/10.1063/1.451630}
  {\bibfield  {journal} {\bibinfo  {journal} {J. Chem. Phys.}\ }\textbf
  {\bibinfo {volume} {85}},\ \bibinfo {pages} {51} (\bibinfo {year}
  {1986})}\BibitemShut {NoStop}%
\bibitem [{\citenamefont {Leopold}\ \emph {et~al.}(1988)\citenamefont
  {Leopold}, \citenamefont {Alml\"{o}f}, \citenamefont {Lineberger},\ and\
  \citenamefont {Taylor}}]{Leopold88}%
  \BibitemOpen
  \bibfield  {author} {\bibinfo {author} {\bibfnamefont {D.~G.}\ \bibnamefont
  {Leopold}}, \bibinfo {author} {\bibfnamefont {J.}~\bibnamefont {Alml\"{o}f}},
  \bibinfo {author} {\bibfnamefont {W.~C.}\ \bibnamefont {Lineberger}}, \ and\
  \bibinfo {author} {\bibfnamefont {P.~R.}\ \bibnamefont {Taylor}},\ }\href
  {\doibase http://dx.doi.org/10.1063/1.453876} {\bibfield  {journal} {\bibinfo
   {journal} {J. Chem. Phys.}\ }\textbf {\bibinfo {volume} {88}},\ \bibinfo
  {pages} {3780} (\bibinfo {year} {1988})}\BibitemShut {NoStop}%
\bibitem [{\citenamefont {{Van Zee}}\ \emph {et~al.}(1992)\citenamefont {{Van
  Zee}}, \citenamefont {Hamrick}, \citenamefont {Li},\ and\ \citenamefont
  {{Weltner Jr.}}}]{VanZee92}%
  \BibitemOpen
  \bibfield  {author} {\bibinfo {author} {\bibfnamefont {R.~J.}\ \bibnamefont
  {{Van Zee}}}, \bibinfo {author} {\bibfnamefont {Y.~M.}\ \bibnamefont
  {Hamrick}}, \bibinfo {author} {\bibfnamefont {S.}~\bibnamefont {Li}}, \ and\
  \bibinfo {author} {\bibfnamefont {W.}~\bibnamefont {{Weltner Jr.}}},\ }\href
  {\doibase http://dx.doi.org/10.1016/0009-2614(92)86138-8} {\bibfield
  {journal} {\bibinfo  {journal} {Chem. Phys. Lett.}\ }\textbf {\bibinfo
  {volume} {195}},\ \bibinfo {pages} {214} (\bibinfo {year}
  {1992})}\BibitemShut {NoStop}%
\bibitem [{\citenamefont {Danset}\ and\ \citenamefont
  {Manceron}(2004)}]{Danset04}%
  \BibitemOpen
  \bibfield  {author} {\bibinfo {author} {\bibfnamefont {D.}~\bibnamefont
  {Danset}}\ and\ \bibinfo {author} {\bibfnamefont {L.}~\bibnamefont
  {Manceron}},\ }\href {\doibase 10.1039/B405149A} {\bibfield  {journal}
  {\bibinfo  {journal} {Phys. Chem. Chem. Phys.}\ }\textbf {\bibinfo {volume}
  {6}},\ \bibinfo {pages} {3928} (\bibinfo {year} {2004})}\BibitemShut
  {NoStop}%
\bibitem [{\citenamefont {H\"ubner}\ and\ \citenamefont
  {Sauer}(2002)}]{Huebner02}%
  \BibitemOpen
  \bibfield  {author} {\bibinfo {author} {\bibfnamefont {O.}~\bibnamefont
  {H\"ubner}}\ and\ \bibinfo {author} {\bibfnamefont {J.}~\bibnamefont
  {Sauer}},\ }\href {\doibase http://dx.doi.org/10.1016/S0009-2614(02)00673-5}
  {\bibfield  {journal} {\bibinfo  {journal} {Chem. Phys. Lett.}\ }\textbf
  {\bibinfo {volume} {358}},\ \bibinfo {pages} {442} (\bibinfo {year}
  {2002})}\BibitemShut {NoStop}%
\bibitem [{\citenamefont {Irigoras}\ \emph {et~al.}(2003)\citenamefont
  {Irigoras}, \citenamefont {del Carmen~Michelini}, \citenamefont {Sicilia},
  \citenamefont {Russo}, \citenamefont {Mercero},\ and\ \citenamefont
  {Ugalde}}]{Irigoras03}%
  \BibitemOpen
  \bibfield  {author} {\bibinfo {author} {\bibfnamefont {A.}~\bibnamefont
  {Irigoras}}, \bibinfo {author} {\bibfnamefont {M.}~\bibnamefont {del
  Carmen~Michelini}}, \bibinfo {author} {\bibfnamefont {E.}~\bibnamefont
  {Sicilia}}, \bibinfo {author} {\bibfnamefont {N.}~\bibnamefont {Russo}},
  \bibinfo {author} {\bibfnamefont {J.~M.}\ \bibnamefont {Mercero}}, \ and\
  \bibinfo {author} {\bibfnamefont {J.~M.}\ \bibnamefont {Ugalde}},\ }\href
  {\doibase http://dx.doi.org/10.1016/S0009-2614(03)00988-6} {\bibfield
  {journal} {\bibinfo  {journal} {Chem. Phys. Lett.}\ }\textbf {\bibinfo
  {volume} {376}},\ \bibinfo {pages} {310} (\bibinfo {year}
  {2003})}\BibitemShut {NoStop}%
\bibitem [{\citenamefont {Wang}, \citenamefont {Khait},\ and\ \citenamefont
  {Hoffmann}(2005)}]{Wang05e}%
  \BibitemOpen
  \bibfield  {author} {\bibinfo {author} {\bibfnamefont {H.}~\bibnamefont
  {Wang}}, \bibinfo {author} {\bibfnamefont {Y.~G.}\ \bibnamefont {Khait}}, \
  and\ \bibinfo {author} {\bibfnamefont {M.~R.}\ \bibnamefont {Hoffmann}},\
  }\href {\doibase 10.1080/00268970512331317327} {\bibfield  {journal}
  {\bibinfo  {journal} {Mol. Phys.}\ }\textbf {\bibinfo {volume} {103}},\
  \bibinfo {pages} {263} (\bibinfo {year} {2005})}\BibitemShut {NoStop}%
\bibitem [{\citenamefont {Rollmann}, \citenamefont {Entel},\ and\ \citenamefont
  {Sahoo}(2006)}]{Rollmann06}%
  \BibitemOpen
  \bibfield  {author} {\bibinfo {author} {\bibfnamefont {G.}~\bibnamefont
  {Rollmann}}, \bibinfo {author} {\bibfnamefont {P.}~\bibnamefont {Entel}}, \
  and\ \bibinfo {author} {\bibfnamefont {S.}~\bibnamefont {Sahoo}},\ }\href
  {\doibase 10.1016/j.commatsci.2004.09.059} {\bibfield  {journal} {\bibinfo
  {journal} {Comput. Mater. Sci.}\ }\textbf {\bibinfo {volume} {35}},\ \bibinfo
  {pages} {275} (\bibinfo {year} {2006})}\BibitemShut {NoStop}%
\bibitem [{\citenamefont {Strandberg}, \citenamefont {Canali},\ and\
  \citenamefont {MacDonald}(2007)}]{Strandberg07}%
  \BibitemOpen
  \bibfield  {author} {\bibinfo {author} {\bibfnamefont {T.~O.}\ \bibnamefont
  {Strandberg}}, \bibinfo {author} {\bibfnamefont {C.~M.}\ \bibnamefont
  {Canali}}, \ and\ \bibinfo {author} {\bibfnamefont {A.~H.}\ \bibnamefont
  {MacDonald}},\ }\href {\doibase 10.1038/nmat1968} {\bibfield  {journal}
  {\bibinfo  {journal} {Nat. Mater.}\ }\textbf {\bibinfo {volume} {6}},\
  \bibinfo {pages} {648} (\bibinfo {year} {2007})}\BibitemShut {NoStop}%
\bibitem [{\citenamefont {Fritsch}\ \emph {et~al.}(2008)\citenamefont
  {Fritsch}, \citenamefont {Koepernik}, \citenamefont {Richter},\ and\
  \citenamefont {Eschrig}}]{Fritsch08}%
  \BibitemOpen
  \bibfield  {author} {\bibinfo {author} {\bibfnamefont {D.}~\bibnamefont
  {Fritsch}}, \bibinfo {author} {\bibfnamefont {K.}~\bibnamefont {Koepernik}},
  \bibinfo {author} {\bibfnamefont {M.}~\bibnamefont {Richter}}, \ and\
  \bibinfo {author} {\bibfnamefont {H.}~\bibnamefont {Eschrig}},\ }\href
  {\doibase 10.1002/jcc.21012} {\bibfield  {journal} {\bibinfo  {journal} {J.
  Comput. Chem.}\ }\textbf {\bibinfo {volume} {29}},\ \bibinfo {pages} {2210}
  (\bibinfo {year} {2008})}\BibitemShut {NoStop}%
\bibitem [{\citenamefont {B{\l}o{\'{n}}ski}\ and\ \citenamefont
  {Hafner}(2009)}]{Blonski09}%
  \BibitemOpen
  \bibfield  {author} {\bibinfo {author} {\bibfnamefont {P.}~\bibnamefont
  {B{\l}o{\'{n}}ski}}\ and\ \bibinfo {author} {\bibfnamefont {J.}~\bibnamefont
  {Hafner}},\ }\href {\doibase 10.1103/PhysRevB.79.224418} {\bibfield
  {journal} {\bibinfo  {journal} {Phys. Rev. B}\ }\textbf {\bibinfo {volume}
  {79}},\ \bibinfo {pages} {224418} (\bibinfo {year} {2009})}\BibitemShut
  {NoStop}%
\bibitem [{\citenamefont {Xiao}\ \emph {et~al.}(2009)\citenamefont {Xiao},
  \citenamefont {Fritsch}, \citenamefont {Kuz'min}, \citenamefont {Koepernik},
  \citenamefont {Eschrig}, \citenamefont {Richter}, \citenamefont {Vietze},\
  and\ \citenamefont {Seifert}}]{Xiao09}%
  \BibitemOpen
  \bibfield  {author} {\bibinfo {author} {\bibfnamefont {R.}~\bibnamefont
  {Xiao}}, \bibinfo {author} {\bibfnamefont {D.}~\bibnamefont {Fritsch}},
  \bibinfo {author} {\bibfnamefont {M.~D.}\ \bibnamefont {Kuz'min}}, \bibinfo
  {author} {\bibfnamefont {K.}~\bibnamefont {Koepernik}}, \bibinfo {author}
  {\bibfnamefont {H.}~\bibnamefont {Eschrig}}, \bibinfo {author} {\bibfnamefont
  {M.}~\bibnamefont {Richter}}, \bibinfo {author} {\bibfnamefont
  {K.}~\bibnamefont {Vietze}}, \ and\ \bibinfo {author} {\bibfnamefont
  {G.}~\bibnamefont {Seifert}},\ }\href {\doibase
  10.1103/PhysRevLett.103.187201} {\bibfield  {journal} {\bibinfo  {journal}
  {Phys. Rev. Lett.}\ }\textbf {\bibinfo {volume} {103}},\ \bibinfo {pages}
  {187201} (\bibinfo {year} {2009})}\BibitemShut {NoStop}%
\bibitem [{\citenamefont {Angeli}\ and\ \citenamefont
  {Cimiraglia}(2011)}]{Angeli11}%
  \BibitemOpen
  \bibfield  {author} {\bibinfo {author} {\bibfnamefont {C.}~\bibnamefont
  {Angeli}}\ and\ \bibinfo {author} {\bibfnamefont {R.}~\bibnamefont
  {Cimiraglia}},\ }\href {\doibase 10.1080/00268976.2011.566586} {\bibfield
  {journal} {\bibinfo  {journal} {Mol. Phys.}\ }\textbf {\bibinfo {volume}
  {109}},\ \bibinfo {pages} {1503} (\bibinfo {year} {2011})}\BibitemShut
  {NoStop}%
\bibitem [{\citenamefont {Hoyer}\ \emph {et~al.}(2014)\citenamefont {Hoyer},
  \citenamefont {Manni}, \citenamefont {Truhlar},\ and\ \citenamefont
  {Gagliardi}}]{Hoyer14}%
  \BibitemOpen
  \bibfield  {author} {\bibinfo {author} {\bibfnamefont {C.~E.}\ \bibnamefont
  {Hoyer}}, \bibinfo {author} {\bibfnamefont {G.~L.}\ \bibnamefont {Manni}},
  \bibinfo {author} {\bibfnamefont {D.~G.}\ \bibnamefont {Truhlar}}, \ and\
  \bibinfo {author} {\bibfnamefont {L.}~\bibnamefont {Gagliardi}},\ }\href
  {\doibase http://dx.doi.org/10.1063/1.4901718} {\bibfield  {journal}
  {\bibinfo  {journal} {J. Chem. Phys.}\ }\textbf {\bibinfo {volume} {141}},\
  \bibinfo {eid} {204309} (\bibinfo {year} {2014})}\BibitemShut {NoStop}%
\bibitem [{\citenamefont {Kalemos}(2015)}]{Kalemos15}%
  \BibitemOpen
  \bibfield  {author} {\bibinfo {author} {\bibfnamefont {A.}~\bibnamefont
  {Kalemos}},\ }\href {\doibase 10.1063/1.4922793} {\bibfield  {journal}
  {\bibinfo  {journal} {J. Chem. Phys.}\ }\textbf {\bibinfo {volume} {142}},\
  \bibinfo {eid} {244304} (\bibinfo {year} {2015})}\BibitemShut {NoStop}%
\bibitem [{\citenamefont {Friedrich}\ and\ \citenamefont
  {Herschbach}(1992)}]{Friedrich92}%
  \BibitemOpen
  \bibfield  {author} {\bibinfo {author} {\bibfnamefont {B.}~\bibnamefont
  {Friedrich}}\ and\ \bibinfo {author} {\bibfnamefont {D.~R.}\ \bibnamefont
  {Herschbach}},\ }\href {\doibase 10.1007/BF01436600} {\bibfield  {journal}
  {\bibinfo  {journal} {Z. Phys. D}\ }\textbf {\bibinfo {volume} {24}},\
  \bibinfo {pages} {25} (\bibinfo {year} {1992})}\BibitemShut {NoStop}%
\bibitem [{\citenamefont {Slenczka}, \citenamefont {Friedrich},\ and\
  \citenamefont {Herschbach}(1994)}]{Slenczka94}%
  \BibitemOpen
  \bibfield  {author} {\bibinfo {author} {\bibfnamefont {A.}~\bibnamefont
  {Slenczka}}, \bibinfo {author} {\bibfnamefont {B.}~\bibnamefont {Friedrich}},
  \ and\ \bibinfo {author} {\bibfnamefont {D.}~\bibnamefont {Herschbach}},\
  }\href {\doibase 10.1103/PhysRevLett.72.1806} {\bibfield  {journal} {\bibinfo
   {journal} {Phys. Rev. Lett.}\ }\textbf {\bibinfo {volume} {72}},\ \bibinfo
  {pages} {1806} (\bibinfo {year} {1994})}\BibitemShut {NoStop}%
\bibitem [{\citenamefont {Loh}\ \emph {et~al.}(1989)\citenamefont {Loh},
  \citenamefont {Hales}, \citenamefont {Lian},\ and\ \citenamefont
  {Armentrout}}]{Loh89}%
  \BibitemOpen
  \bibfield  {author} {\bibinfo {author} {\bibfnamefont {S.~K.}\ \bibnamefont
  {Loh}}, \bibinfo {author} {\bibfnamefont {D.~A.}\ \bibnamefont {Hales}},
  \bibinfo {author} {\bibfnamefont {L.}~\bibnamefont {Lian}}, \ and\ \bibinfo
  {author} {\bibfnamefont {P.~B.}\ \bibnamefont {Armentrout}},\ }\href
  {\doibase 10.1063/1.456452} {\bibfield  {journal} {\bibinfo  {journal} {J.
  Chem. Phys.}\ }\textbf {\bibinfo {volume} {90}},\ \bibinfo {pages} {5466}
  (\bibinfo {year} {1989})}\BibitemShut {NoStop}%
\bibitem [{\citenamefont {Lian}, \citenamefont {Su},\ and\ \citenamefont
  {Armentrout}(1992)}]{Lian92a}%
  \BibitemOpen
  \bibfield  {author} {\bibinfo {author} {\bibfnamefont {L.}~\bibnamefont
  {Lian}}, \bibinfo {author} {\bibfnamefont {C.-X.}\ \bibnamefont {Su}}, \ and\
  \bibinfo {author} {\bibfnamefont {P.~B.}\ \bibnamefont {Armentrout}},\ }\href
  {\doibase 10.1063/1.463912} {\bibfield  {journal} {\bibinfo  {journal} {J.
  Chem. Phys.}\ }\textbf {\bibinfo {volume} {97}},\ \bibinfo {pages} {4072}
  (\bibinfo {year} {1992})}\BibitemShut {NoStop}%
\bibitem [{\citenamefont {Russon}\ \emph {et~al.}(1993)\citenamefont {Russon},
  \citenamefont {Heidecke}, \citenamefont {Birke}, \citenamefont {Conceicao},
  \citenamefont {Armentrout},\ and\ \citenamefont {Morse}}]{Russon93}%
  \BibitemOpen
  \bibfield  {author} {\bibinfo {author} {\bibfnamefont {L.~M.}\ \bibnamefont
  {Russon}}, \bibinfo {author} {\bibfnamefont {S.~A.}\ \bibnamefont
  {Heidecke}}, \bibinfo {author} {\bibfnamefont {M.~K.}\ \bibnamefont {Birke}},
  \bibinfo {author} {\bibfnamefont {J.}~\bibnamefont {Conceicao}}, \bibinfo
  {author} {\bibfnamefont {P.}~\bibnamefont {Armentrout}}, \ and\ \bibinfo
  {author} {\bibfnamefont {M.~D.}\ \bibnamefont {Morse}},\ }\href {\doibase
  10.1016/0009-2614(93)90002-I} {\bibfield  {journal} {\bibinfo  {journal}
  {Chem. Phys. Lett.}\ }\textbf {\bibinfo {volume} {204}},\ \bibinfo {pages}
  {235} (\bibinfo {year} {1993})}\BibitemShut {NoStop}%
\bibitem [{\citenamefont {Russon}\ \emph {et~al.}(1994)\citenamefont {Russon},
  \citenamefont {Heidecke}, \citenamefont {Birke}, \citenamefont {Conceicao},
  \citenamefont {Morse},\ and\ \citenamefont {Armentrout}}]{Russon94}%
  \BibitemOpen
  \bibfield  {author} {\bibinfo {author} {\bibfnamefont {L.~M.}\ \bibnamefont
  {Russon}}, \bibinfo {author} {\bibfnamefont {S.~A.}\ \bibnamefont
  {Heidecke}}, \bibinfo {author} {\bibfnamefont {M.~K.}\ \bibnamefont {Birke}},
  \bibinfo {author} {\bibfnamefont {J.}~\bibnamefont {Conceicao}}, \bibinfo
  {author} {\bibfnamefont {M.~D.}\ \bibnamefont {Morse}}, \ and\ \bibinfo
  {author} {\bibfnamefont {P.~B.}\ \bibnamefont {Armentrout}},\ }\href
  {\doibase 10.1063/1.466265} {\bibfield  {journal} {\bibinfo  {journal} {J.
  Chem. Phys.}\ }\textbf {\bibinfo {volume} {100}},\ \bibinfo {pages} {4747}
  (\bibinfo {year} {1994})}\BibitemShut {NoStop}%
\bibitem [{\citenamefont {Hales}\ \emph {et~al.}(1994)\citenamefont {Hales},
  \citenamefont {Su}, \citenamefont {Lian},\ and\ \citenamefont
  {Armentrout}}]{Hales94}%
  \BibitemOpen
  \bibfield  {author} {\bibinfo {author} {\bibfnamefont {D.~A.}\ \bibnamefont
  {Hales}}, \bibinfo {author} {\bibfnamefont {C.-X.}\ \bibnamefont {Su}},
  \bibinfo {author} {\bibfnamefont {L.}~\bibnamefont {Lian}}, \ and\ \bibinfo
  {author} {\bibfnamefont {P.~B.}\ \bibnamefont {Armentrout}},\ }\href
  {\doibase 10.1063/1.466636} {\bibfield  {journal} {\bibinfo  {journal} {J.
  Chem. Phys.}\ }\textbf {\bibinfo {volume} {100}},\ \bibinfo {pages} {1049}
  (\bibinfo {year} {1994})}\BibitemShut {NoStop}%
\bibitem [{\citenamefont {Rollmann}, \citenamefont {Herper},\ and\
  \citenamefont {Entel}(2006)}]{Rollmann06a}%
  \BibitemOpen
  \bibfield  {author} {\bibinfo {author} {\bibfnamefont {G.}~\bibnamefont
  {Rollmann}}, \bibinfo {author} {\bibfnamefont {H.~C.}\ \bibnamefont
  {Herper}}, \ and\ \bibinfo {author} {\bibfnamefont {P.}~\bibnamefont
  {Entel}},\ }\href {\doibase 10.1021/jp061794s} {\bibfield  {journal}
  {\bibinfo  {journal} {J. Phys. Chem. A}\ }\textbf {\bibinfo {volume} {110}},\
  \bibinfo {pages} {10799} (\bibinfo {year} {2006})}\BibitemShut {NoStop}%
\bibitem [{\citenamefont {Chiodo}\ \emph {et~al.}(2006)\citenamefont {Chiodo},
  \citenamefont {Rivalta}, \citenamefont {Michelini}, \citenamefont {Russo},
  \citenamefont {Sicilia},\ and\ \citenamefont {Ugalde}}]{Chiodo06}%
  \BibitemOpen
  \bibfield  {author} {\bibinfo {author} {\bibfnamefont {S.}~\bibnamefont
  {Chiodo}}, \bibinfo {author} {\bibfnamefont {I.}~\bibnamefont {Rivalta}},
  \bibinfo {author} {\bibfnamefont {M.~d.~C.}\ \bibnamefont {Michelini}},
  \bibinfo {author} {\bibfnamefont {N.}~\bibnamefont {Russo}}, \bibinfo
  {author} {\bibfnamefont {E.}~\bibnamefont {Sicilia}}, \ and\ \bibinfo
  {author} {\bibfnamefont {J.~M.}\ \bibnamefont {Ugalde}},\ }\href {\doibase
  10.1021/jp064611a} {\bibfield  {journal} {\bibinfo  {journal} {J. Phys. Chem.
  A}\ }\textbf {\bibinfo {volume} {110}},\ \bibinfo {pages} {12501} (\bibinfo
  {year} {2006})}\BibitemShut {NoStop}%
\bibitem [{\citenamefont {Jamorski}\ \emph {et~al.}(1997)\citenamefont
  {Jamorski}, \citenamefont {Martinez}, \citenamefont {Castro},\ and\
  \citenamefont {Salahub}}]{Jamorski97}%
  \BibitemOpen
  \bibfield  {author} {\bibinfo {author} {\bibfnamefont {C.}~\bibnamefont
  {Jamorski}}, \bibinfo {author} {\bibfnamefont {A.}~\bibnamefont {Martinez}},
  \bibinfo {author} {\bibfnamefont {M.}~\bibnamefont {Castro}}, \ and\ \bibinfo
  {author} {\bibfnamefont {D.~R.}\ \bibnamefont {Salahub}},\ }\href {\doibase
  10.1103/PhysRevB.55.10905} {\bibfield  {journal} {\bibinfo  {journal} {Phys.
  Rev. B}\ }\textbf {\bibinfo {volume} {55}},\ \bibinfo {pages} {10905}
  (\bibinfo {year} {1997})}\BibitemShut {NoStop}%
\bibitem [{\citenamefont {Peredkov}\ \emph {et~al.}(2011)\citenamefont
  {Peredkov}, \citenamefont {Neeb}, \citenamefont {Eberhardt}, \citenamefont
  {Meyer}, \citenamefont {Tombers}, \citenamefont {Kampschulte},\ and\
  \citenamefont {Niedner-Schatteburg}}]{Peredkov11b}%
  \BibitemOpen
  \bibfield  {author} {\bibinfo {author} {\bibfnamefont {S.}~\bibnamefont
  {Peredkov}}, \bibinfo {author} {\bibfnamefont {M.}~\bibnamefont {Neeb}},
  \bibinfo {author} {\bibfnamefont {W.}~\bibnamefont {Eberhardt}}, \bibinfo
  {author} {\bibfnamefont {J.}~\bibnamefont {Meyer}}, \bibinfo {author}
  {\bibfnamefont {M.}~\bibnamefont {Tombers}}, \bibinfo {author} {\bibfnamefont
  {H.}~\bibnamefont {Kampschulte}}, \ and\ \bibinfo {author} {\bibfnamefont
  {G.}~\bibnamefont {Niedner-Schatteburg}},\ }\href {\doibase
  10.1103/PhysRevLett.107.233401} {\bibfield  {journal} {\bibinfo  {journal}
  {Phys. Rev. Lett.}\ }\textbf {\bibinfo {volume} {107}},\ \bibinfo {pages}
  {233401} (\bibinfo {year} {2011})}\BibitemShut {NoStop}%
\bibitem [{\citenamefont {Niemeyer}\ \emph {et~al.}(2012)\citenamefont
  {Niemeyer}, \citenamefont {Hirsch}, \citenamefont {Zamudio-Bayer},
  \citenamefont {Langenberg}, \citenamefont {Vogel}, \citenamefont {Kossick},
  \citenamefont {Ebrecht}, \citenamefont {Egashira}, \citenamefont {Terasaki},
  \citenamefont {M\"oller}, \citenamefont {v.~Issendorff},\ and\ \citenamefont
  {Lau}}]{Niemeyer12}%
  \BibitemOpen
  \bibfield  {author} {\bibinfo {author} {\bibfnamefont {M.}~\bibnamefont
  {Niemeyer}}, \bibinfo {author} {\bibfnamefont {K.}~\bibnamefont {Hirsch}},
  \bibinfo {author} {\bibfnamefont {V.}~\bibnamefont {Zamudio-Bayer}}, \bibinfo
  {author} {\bibfnamefont {A.}~\bibnamefont {Langenberg}}, \bibinfo {author}
  {\bibfnamefont {M.}~\bibnamefont {Vogel}}, \bibinfo {author} {\bibfnamefont
  {M.}~\bibnamefont {Kossick}}, \bibinfo {author} {\bibfnamefont
  {C.}~\bibnamefont {Ebrecht}}, \bibinfo {author} {\bibfnamefont
  {K.}~\bibnamefont {Egashira}}, \bibinfo {author} {\bibfnamefont
  {A.}~\bibnamefont {Terasaki}}, \bibinfo {author} {\bibfnamefont
  {T.}~\bibnamefont {M\"oller}}, \bibinfo {author} {\bibfnamefont
  {B.}~\bibnamefont {v.~Issendorff}}, \ and\ \bibinfo {author} {\bibfnamefont
  {J.~T.}\ \bibnamefont {Lau}},\ }\href {\doibase
  10.1103/PhysRevLett.108.057201} {\bibfield  {journal} {\bibinfo  {journal}
  {Phys. Rev. Lett.}\ }\textbf {\bibinfo {volume} {108}},\ \bibinfo {pages}
  {057201} (\bibinfo {year} {2012})}\BibitemShut {NoStop}%
\bibitem [{\citenamefont {Zamudio-Bayer}\ \emph {et~al.}(2013)\citenamefont
  {Zamudio-Bayer}, \citenamefont {Leppert}, \citenamefont {Hirsch},
  \citenamefont {Langenberg}, \citenamefont {Rittmann}, \citenamefont
  {Kossick}, \citenamefont {Vogel}, \citenamefont {Richter}, \citenamefont
  {Terasaki}, \citenamefont {M\"oller}, \citenamefont {v.~Issendorff},
  \citenamefont {K\"ummel},\ and\ \citenamefont {Lau}}]{ZamudioBayer13}%
  \BibitemOpen
  \bibfield  {author} {\bibinfo {author} {\bibfnamefont {V.}~\bibnamefont
  {Zamudio-Bayer}}, \bibinfo {author} {\bibfnamefont {L.}~\bibnamefont
  {Leppert}}, \bibinfo {author} {\bibfnamefont {K.}~\bibnamefont {Hirsch}},
  \bibinfo {author} {\bibfnamefont {A.}~\bibnamefont {Langenberg}}, \bibinfo
  {author} {\bibfnamefont {J.}~\bibnamefont {Rittmann}}, \bibinfo {author}
  {\bibfnamefont {M.}~\bibnamefont {Kossick}}, \bibinfo {author} {\bibfnamefont
  {M.}~\bibnamefont {Vogel}}, \bibinfo {author} {\bibfnamefont
  {R.}~\bibnamefont {Richter}}, \bibinfo {author} {\bibfnamefont
  {A.}~\bibnamefont {Terasaki}}, \bibinfo {author} {\bibfnamefont
  {T.}~\bibnamefont {M\"oller}}, \bibinfo {author} {\bibfnamefont
  {B.}~\bibnamefont {v.~Issendorff}}, \bibinfo {author} {\bibfnamefont
  {S.}~\bibnamefont {K\"ummel}}, \ and\ \bibinfo {author} {\bibfnamefont
  {J.~T.}\ \bibnamefont {Lau}},\ }\href {\doibase 10.1103/PhysRevB.88.115425}
  {\bibfield  {journal} {\bibinfo  {journal} {Phys. Rev. B}\ }\textbf {\bibinfo
  {volume} {88}},\ \bibinfo {pages} {115425} (\bibinfo {year}
  {2013})}\BibitemShut {NoStop}%
\bibitem [{\citenamefont {Langenberg}\ \emph {et~al.}(2014)\citenamefont
  {Langenberg}, \citenamefont {Hirsch}, \citenamefont {{\L}awicki},
  \citenamefont {Zamudio-Bayer}, \citenamefont {Niemeyer}, \citenamefont
  {Chmiela}, \citenamefont {Langbehn}, \citenamefont {Terasaki}, \citenamefont
  {Issendorff},\ and\ \citenamefont {Lau}}]{Langenberg14}%
  \BibitemOpen
  \bibfield  {author} {\bibinfo {author} {\bibfnamefont {A.}~\bibnamefont
  {Langenberg}}, \bibinfo {author} {\bibfnamefont {K.}~\bibnamefont {Hirsch}},
  \bibinfo {author} {\bibfnamefont {A.}~\bibnamefont {{\L}awicki}}, \bibinfo
  {author} {\bibfnamefont {V.}~\bibnamefont {Zamudio-Bayer}}, \bibinfo {author}
  {\bibfnamefont {M.}~\bibnamefont {Niemeyer}}, \bibinfo {author}
  {\bibfnamefont {P.}~\bibnamefont {Chmiela}}, \bibinfo {author} {\bibfnamefont
  {B.}~\bibnamefont {Langbehn}}, \bibinfo {author} {\bibfnamefont
  {A.}~\bibnamefont {Terasaki}}, \bibinfo {author} {\bibfnamefont {B.~v.}\
  \bibnamefont {Issendorff}}, \ and\ \bibinfo {author} {\bibfnamefont {J.~T.}\
  \bibnamefont {Lau}},\ }\href {\doibase 10.1103/PhysRevB.90.184420} {\bibfield
   {journal} {\bibinfo  {journal} {Phys. Rev. B}\ }\textbf {\bibinfo {volume}
  {90}},\ \bibinfo {pages} {184420} (\bibinfo {year} {2014})}\BibitemShut
  {NoStop}%
\bibitem [{\citenamefont {Hirsch}\ \emph {et~al.}(2015)\citenamefont {Hirsch},
  \citenamefont {Zamudio-Bayer}, \citenamefont {Langenberg}, \citenamefont
  {Niemeyer}, \citenamefont {Langbehn}, \citenamefont {M\"oller}, \citenamefont
  {Terasaki}, \citenamefont {Issendorff},\ and\ \citenamefont
  {Lau}}]{Hirsch15a}%
  \BibitemOpen
  \bibfield  {author} {\bibinfo {author} {\bibfnamefont {K.}~\bibnamefont
  {Hirsch}}, \bibinfo {author} {\bibfnamefont {V.}~\bibnamefont
  {Zamudio-Bayer}}, \bibinfo {author} {\bibfnamefont {A.}~\bibnamefont
  {Langenberg}}, \bibinfo {author} {\bibfnamefont {M.}~\bibnamefont
  {Niemeyer}}, \bibinfo {author} {\bibfnamefont {B.}~\bibnamefont {Langbehn}},
  \bibinfo {author} {\bibfnamefont {T.}~\bibnamefont {M\"oller}}, \bibinfo
  {author} {\bibfnamefont {A.}~\bibnamefont {Terasaki}}, \bibinfo {author}
  {\bibfnamefont {B.~v.}\ \bibnamefont {Issendorff}}, \ and\ \bibinfo {author}
  {\bibfnamefont {J.~T.}\ \bibnamefont {Lau}},\ }\href {\doibase
  10.1103/PhysRevLett.114.087202} {\bibfield  {journal} {\bibinfo  {journal}
  {Phys. Rev. Lett.}\ }\textbf {\bibinfo {volume} {114}},\ \bibinfo {pages}
  {087202} (\bibinfo {year} {2015})}\BibitemShut {NoStop}%
\bibitem [{\citenamefont {Zamudio-Bayer}\ \emph
  {et~al.}(2015{\natexlab{a}})\citenamefont {Zamudio-Bayer}, \citenamefont
  {Hirsch}, \citenamefont {Langenberg}, \citenamefont {Niemeyer}, \citenamefont
  {Vogel}, \citenamefont {{\L}awicki}, \citenamefont {Terasaki}, \citenamefont
  {Lau},\ and\ \citenamefont {von Issendorff}}]{ZamudioBayer15a}%
  \BibitemOpen
  \bibfield  {author} {\bibinfo {author} {\bibfnamefont {V.}~\bibnamefont
  {Zamudio-Bayer}}, \bibinfo {author} {\bibfnamefont {K.}~\bibnamefont
  {Hirsch}}, \bibinfo {author} {\bibfnamefont {A.}~\bibnamefont {Langenberg}},
  \bibinfo {author} {\bibfnamefont {M.}~\bibnamefont {Niemeyer}}, \bibinfo
  {author} {\bibfnamefont {M.}~\bibnamefont {Vogel}}, \bibinfo {author}
  {\bibfnamefont {A.}~\bibnamefont {{\L}awicki}}, \bibinfo {author}
  {\bibfnamefont {A.}~\bibnamefont {Terasaki}}, \bibinfo {author}
  {\bibfnamefont {J.~T.}\ \bibnamefont {Lau}}, \ and\ \bibinfo {author}
  {\bibfnamefont {B.}~\bibnamefont {von Issendorff}},\ }\href {\doibase
  10.1002/anie.201411018} {\bibfield  {journal} {\bibinfo  {journal} {Angew.
  Chem. Int. Ed.}\ }\textbf {\bibinfo {volume} {54}},\ \bibinfo {pages} {4498}
  (\bibinfo {year} {2015}{\natexlab{a}})}\BibitemShut {NoStop}%
\bibitem [{\citenamefont {Zamudio-Bayer}\ \emph
  {et~al.}(2015{\natexlab{b}})\citenamefont {Zamudio-Bayer}, \citenamefont
  {Hirsch}, \citenamefont {Langenberg}, \citenamefont {Kossick}, \citenamefont
  {{\L}awicki}, \citenamefont {Terasaki}, \citenamefont {v.~Issendorff},\ and\
  \citenamefont {Lau}}]{ZamudioBayer15b}%
  \BibitemOpen
  \bibfield  {author} {\bibinfo {author} {\bibfnamefont {V.}~\bibnamefont
  {Zamudio-Bayer}}, \bibinfo {author} {\bibfnamefont {K.}~\bibnamefont
  {Hirsch}}, \bibinfo {author} {\bibfnamefont {A.}~\bibnamefont {Langenberg}},
  \bibinfo {author} {\bibfnamefont {M.}~\bibnamefont {Kossick}}, \bibinfo
  {author} {\bibfnamefont {A.}~\bibnamefont {{\L}awicki}}, \bibinfo {author}
  {\bibfnamefont {A.}~\bibnamefont {Terasaki}}, \bibinfo {author}
  {\bibfnamefont {B.}~\bibnamefont {v.~Issendorff}}, \ and\ \bibinfo {author}
  {\bibfnamefont {J.~T.}\ \bibnamefont {Lau}},\ }\href {\doibase
  http://dx.doi.org/10.1063/1.4922487} {\bibfield  {journal} {\bibinfo
  {journal} {J. Chem. Phys.}\ }\textbf {\bibinfo {volume} {142}},\ \bibinfo
  {pages} {234301} (\bibinfo {year} {2015}{\natexlab{b}})}\BibitemShut
  {NoStop}%
\bibitem [{\citenamefont {Hirsch}\ \emph {et~al.}(2009)\citenamefont {Hirsch},
  \citenamefont {Lau}, \citenamefont {Klar}, \citenamefont {Langenberg},
  \citenamefont {Probst}, \citenamefont {Rittmann}, \citenamefont {Vogel},
  \citenamefont {Zamudio-Bayer}, \citenamefont {M\"oller},\ and\ \citenamefont
  {von Issendorff}}]{Hirsch09}%
  \BibitemOpen
  \bibfield  {author} {\bibinfo {author} {\bibfnamefont {K.}~\bibnamefont
  {Hirsch}}, \bibinfo {author} {\bibfnamefont {J.~T.}\ \bibnamefont {Lau}},
  \bibinfo {author} {\bibfnamefont {P.}~\bibnamefont {Klar}}, \bibinfo {author}
  {\bibfnamefont {A.}~\bibnamefont {Langenberg}}, \bibinfo {author}
  {\bibfnamefont {J.}~\bibnamefont {Probst}}, \bibinfo {author} {\bibfnamefont
  {J.}~\bibnamefont {Rittmann}}, \bibinfo {author} {\bibfnamefont
  {M.}~\bibnamefont {Vogel}}, \bibinfo {author} {\bibfnamefont
  {V.}~\bibnamefont {Zamudio-Bayer}}, \bibinfo {author} {\bibfnamefont
  {T.}~\bibnamefont {M\"oller}}, \ and\ \bibinfo {author} {\bibfnamefont
  {B.}~\bibnamefont {von Issendorff}},\ }\href {\doibase
  http://dx.doi.org/10.1088/0953-4075/42/15/154029} {\bibfield  {journal}
  {\bibinfo  {journal} {J. Phys. B: At. Mol. Opt. Phys.}\ }\textbf {\bibinfo
  {volume} {42}},\ \bibinfo {pages} {154029} (\bibinfo {year}
  {2009})}\BibitemShut {NoStop}%
\bibitem [{\citenamefont {Major}\ and\ \citenamefont
  {Dehmelt}(1968)}]{Major68}%
  \BibitemOpen
  \bibfield  {author} {\bibinfo {author} {\bibfnamefont {F.~G.}\ \bibnamefont
  {Major}}\ and\ \bibinfo {author} {\bibfnamefont {H.~G.}\ \bibnamefont
  {Dehmelt}},\ }\href {\doibase 10.1103/PhysRev.170.91} {\bibfield  {journal}
  {\bibinfo  {journal} {Phys. Rev.}\ }\textbf {\bibinfo {volume} {170}},\
  \bibinfo {pages} {91} (\bibinfo {year} {1968})}\BibitemShut {NoStop}%
\bibitem [{\citenamefont {Hirsch}\ \emph
  {et~al.}(2012{\natexlab{a}})\citenamefont {Hirsch}, \citenamefont
  {Zamudio-Bayer}, \citenamefont {Ameseder}, \citenamefont {Langenberg},
  \citenamefont {Rittmann}, \citenamefont {Vogel}, \citenamefont {M\"oller},
  \citenamefont {v.~Issendorff},\ and\ \citenamefont {Lau}}]{Hirsch12a}%
  \BibitemOpen
  \bibfield  {author} {\bibinfo {author} {\bibfnamefont {K.}~\bibnamefont
  {Hirsch}}, \bibinfo {author} {\bibfnamefont {V.}~\bibnamefont
  {Zamudio-Bayer}}, \bibinfo {author} {\bibfnamefont {F.}~\bibnamefont
  {Ameseder}}, \bibinfo {author} {\bibfnamefont {A.}~\bibnamefont
  {Langenberg}}, \bibinfo {author} {\bibfnamefont {J.}~\bibnamefont
  {Rittmann}}, \bibinfo {author} {\bibfnamefont {M.}~\bibnamefont {Vogel}},
  \bibinfo {author} {\bibfnamefont {T.}~\bibnamefont {M\"oller}}, \bibinfo
  {author} {\bibfnamefont {B.}~\bibnamefont {v.~Issendorff}}, \ and\ \bibinfo
  {author} {\bibfnamefont {J.~T.}\ \bibnamefont {Lau}},\ }\href {\doibase
  10.1103/PhysRevA.85.062501} {\bibfield  {journal} {\bibinfo  {journal} {Phys.
  Rev. A}\ }\textbf {\bibinfo {volume} {85}},\ \bibinfo {pages} {062501}
  (\bibinfo {year} {2012}{\natexlab{a}})}\BibitemShut {NoStop}%
\bibitem [{\citenamefont {Terasaki}, \citenamefont {Majima},\ and\
  \citenamefont {Kondow}(2007)}]{Terasaki07}%
  \BibitemOpen
  \bibfield  {author} {\bibinfo {author} {\bibfnamefont {A.}~\bibnamefont
  {Terasaki}}, \bibinfo {author} {\bibfnamefont {T.}~\bibnamefont {Majima}}, \
  and\ \bibinfo {author} {\bibfnamefont {T.}~\bibnamefont {Kondow}},\ }\href
  {\doibase 10.1063/1.2822022} {\bibfield  {journal} {\bibinfo  {journal} {J.
  Chem. Phys.}\ }\textbf {\bibinfo {volume} {127}},\ \bibinfo {eid} {231101}
  (\bibinfo {year} {2007})}\BibitemShut {NoStop}%
\bibitem [{\citenamefont {Thole}\ \emph {et~al.}(1992)\citenamefont {Thole},
  \citenamefont {Carra}, \citenamefont {Sette},\ and\ \citenamefont {van~der
  Laan}}]{Thole92}%
  \BibitemOpen
  \bibfield  {author} {\bibinfo {author} {\bibfnamefont {B.~T.}\ \bibnamefont
  {Thole}}, \bibinfo {author} {\bibfnamefont {P.}~\bibnamefont {Carra}},
  \bibinfo {author} {\bibfnamefont {F.}~\bibnamefont {Sette}}, \ and\ \bibinfo
  {author} {\bibfnamefont {G.}~\bibnamefont {van~der Laan}},\ }\href {\doibase
  10.1103/PhysRevLett.68.1943} {\bibfield  {journal} {\bibinfo  {journal}
  {Phys. Rev. Lett.}\ }\textbf {\bibinfo {volume} {68}},\ \bibinfo {pages}
  {1943} (\bibinfo {year} {1992})}\BibitemShut {NoStop}%
\bibitem [{\citenamefont {Carra}\ \emph {et~al.}(1993)\citenamefont {Carra},
  \citenamefont {Thole}, \citenamefont {Altarelli},\ and\ \citenamefont
  {Wang}}]{Carra93}%
  \BibitemOpen
  \bibfield  {author} {\bibinfo {author} {\bibfnamefont {P.}~\bibnamefont
  {Carra}}, \bibinfo {author} {\bibfnamefont {B.~T.}\ \bibnamefont {Thole}},
  \bibinfo {author} {\bibfnamefont {M.}~\bibnamefont {Altarelli}}, \ and\
  \bibinfo {author} {\bibfnamefont {X.}~\bibnamefont {Wang}},\ }\href {\doibase
  10.1103/PhysRevLett.70.694} {\bibfield  {journal} {\bibinfo  {journal} {Phys.
  Rev. Lett.}\ }\textbf {\bibinfo {volume} {70}},\ \bibinfo {pages} {694}
  (\bibinfo {year} {1993})}\BibitemShut {NoStop}%
\bibitem [{\citenamefont {Wu}\ and\ \citenamefont {Freeman}(1994)}]{Wu94}%
  \BibitemOpen
  \bibfield  {author} {\bibinfo {author} {\bibfnamefont {R.}~\bibnamefont
  {Wu}}\ and\ \bibinfo {author} {\bibfnamefont {A.~J.}\ \bibnamefont
  {Freeman}},\ }\href {\doibase 10.1103/PhysRevLett.73.1994} {\bibfield
  {journal} {\bibinfo  {journal} {Phys. Rev. Lett.}\ }\textbf {\bibinfo
  {volume} {73}},\ \bibinfo {pages} {1994} (\bibinfo {year}
  {1994})}\BibitemShut {NoStop}%
\bibitem [{\citenamefont {St\"ohr}(1995)}]{Stoehr95a}%
  \BibitemOpen
  \bibfield  {author} {\bibinfo {author} {\bibfnamefont {J.}~\bibnamefont
  {St\"ohr}},\ }\href@noop {} {\bibfield  {journal} {\bibinfo  {journal} {J.
  Electron Spectrosc. Relat. Phenom.}\ }\textbf {\bibinfo {volume} {75}},\
  \bibinfo {pages} {253} (\bibinfo {year} {1995})}\BibitemShut {NoStop}%
\bibitem [{\citenamefont {St\"ohr}\ and\ \citenamefont
  {K\"onig}(1995)}]{Stoehr95b}%
  \BibitemOpen
  \bibfield  {author} {\bibinfo {author} {\bibfnamefont {J.}~\bibnamefont
  {St\"ohr}}\ and\ \bibinfo {author} {\bibfnamefont {H.}~\bibnamefont
  {K\"onig}},\ }\href {\doibase 10.1103/PhysRevLett.75.3748} {\bibfield
  {journal} {\bibinfo  {journal} {Phys. Rev. Lett.}\ }\textbf {\bibinfo
  {volume} {75}},\ \bibinfo {pages} {3748} (\bibinfo {year}
  {1995})}\BibitemShut {NoStop}%
\bibitem [{\citenamefont {D\"urr}\ and\ \citenamefont {van~der
  Laan}(1996)}]{Duerr96a}%
  \BibitemOpen
  \bibfield  {author} {\bibinfo {author} {\bibfnamefont {H.~A.}\ \bibnamefont
  {D\"urr}}\ and\ \bibinfo {author} {\bibfnamefont {G.}~\bibnamefont {van~der
  Laan}},\ }\href {\doibase 10.1103/PhysRevB.54.R760} {\bibfield  {journal}
  {\bibinfo  {journal} {Phys. Rev. B}\ }\textbf {\bibinfo {volume} {54}},\
  \bibinfo {pages} {R760} (\bibinfo {year} {1996})}\BibitemShut {NoStop}%
\bibitem [{\citenamefont {van~der Laan}\ and\ \citenamefont
  {Figueroa}(2014)}]{vanderLaan14}%
  \BibitemOpen
  \bibfield  {author} {\bibinfo {author} {\bibfnamefont {G.}~\bibnamefont
  {van~der Laan}}\ and\ \bibinfo {author} {\bibfnamefont {A.~I.}\ \bibnamefont
  {Figueroa}},\ }\href {\doibase http://dx.doi.org/10.1016/j.ccr.2014.03.018}
  {\bibfield  {journal} {\bibinfo  {journal} {Coord. Chem. Rev.}\ }\textbf
  {\bibinfo {volume} {277--278}},\ \bibinfo {pages} {95} (\bibinfo {year}
  {2014})}\BibitemShut {NoStop}%
\bibitem [{\citenamefont {Hirsch}\ \emph
  {et~al.}(2012{\natexlab{b}})\citenamefont {Hirsch}, \citenamefont
  {Zamudio-Bayer}, \citenamefont {Rittmann}, \citenamefont {Langenberg},
  \citenamefont {Vogel}, \citenamefont {M\"oller}, \citenamefont
  {v.~Issendorff},\ and\ \citenamefont {Lau}}]{Hirsch12b}%
  \BibitemOpen
  \bibfield  {author} {\bibinfo {author} {\bibfnamefont {K.}~\bibnamefont
  {Hirsch}}, \bibinfo {author} {\bibfnamefont {V.}~\bibnamefont
  {Zamudio-Bayer}}, \bibinfo {author} {\bibfnamefont {J.}~\bibnamefont
  {Rittmann}}, \bibinfo {author} {\bibfnamefont {A.}~\bibnamefont
  {Langenberg}}, \bibinfo {author} {\bibfnamefont {M.}~\bibnamefont {Vogel}},
  \bibinfo {author} {\bibfnamefont {T.}~\bibnamefont {M\"oller}}, \bibinfo
  {author} {\bibfnamefont {B.}~\bibnamefont {v.~Issendorff}}, \ and\ \bibinfo
  {author} {\bibfnamefont {J.~T.}\ \bibnamefont {Lau}},\ }\href {\doibase
  10.1103/PhysRevB.86.165402} {\bibfield  {journal} {\bibinfo  {journal} {Phys.
  Rev. B}\ }\textbf {\bibinfo {volume} {86}},\ \bibinfo {pages} {165402}
  (\bibinfo {year} {2012}{\natexlab{b}})}\BibitemShut {NoStop}%
\bibitem [{\citenamefont {de~Groot}(2005)}]{deGroot05}%
  \BibitemOpen
  \bibfield  {author} {\bibinfo {author} {\bibfnamefont {F.}~\bibnamefont
  {de~Groot}},\ }\href@noop {} {\bibfield  {journal} {\bibinfo  {journal}
  {Coord. Chem. Rev.}\ }\textbf {\bibinfo {volume} {249}},\ \bibinfo {pages}
  {31} (\bibinfo {year} {2005})}\BibitemShut {NoStop}%
\bibitem [{\citenamefont {Martins}\ \emph {et~al.}(2003)\citenamefont
  {Martins}, \citenamefont {Godehusen}, \citenamefont {Richter},\ and\
  \citenamefont {Zimmermann}}]{Martins03}%
  \BibitemOpen
  \bibfield  {author} {\bibinfo {author} {\bibfnamefont {M.}~\bibnamefont
  {Martins}}, \bibinfo {author} {\bibfnamefont {K.}~\bibnamefont {Godehusen}},
  \bibinfo {author} {\bibfnamefont {T.}~\bibnamefont {Richter}}, \ and\
  \bibinfo {author} {\bibfnamefont {P.}~\bibnamefont {Zimmermann}},\ }\href
  {\doibase 10.1063/1.1536372} {\bibfield  {journal} {\bibinfo  {journal} {AIP
  Conf. Proc.}\ }\textbf {\bibinfo {volume} {652}},\ \bibinfo {pages} {153}
  (\bibinfo {year} {2003})}\BibitemShut {NoStop}%
\bibitem [{\citenamefont {Richter}\ \emph {et~al.}(2004)\citenamefont
  {Richter}, \citenamefont {Godehusen}, \citenamefont {Martins}, \citenamefont
  {Wolff},\ and\ \citenamefont {Zimmermann}}]{Richter04a}%
  \BibitemOpen
  \bibfield  {author} {\bibinfo {author} {\bibfnamefont {T.}~\bibnamefont
  {Richter}}, \bibinfo {author} {\bibfnamefont {K.}~\bibnamefont {Godehusen}},
  \bibinfo {author} {\bibfnamefont {M.}~\bibnamefont {Martins}}, \bibinfo
  {author} {\bibfnamefont {T.}~\bibnamefont {Wolff}}, \ and\ \bibinfo {author}
  {\bibfnamefont {P.}~\bibnamefont {Zimmermann}},\ }\href
  {http://link.aps.org/abstract/PRL/v93/e023002} {\bibfield  {journal}
  {\bibinfo  {journal} {Phys. Rev. Lett.}\ }\textbf {\bibinfo {volume} {93}},\
  \bibinfo {eid} {023002} (\bibinfo {year} {2004})}\BibitemShut {NoStop}%
\bibitem [{\citenamefont {Loh}\ \emph {et~al.}(1988)\citenamefont {Loh},
  \citenamefont {Lian}, \citenamefont {Hales},\ and\ \citenamefont
  {Armentrout}}]{Loh88}%
  \BibitemOpen
  \bibfield  {author} {\bibinfo {author} {\bibfnamefont {S.~K.}\ \bibnamefont
  {Loh}}, \bibinfo {author} {\bibfnamefont {L.}~\bibnamefont {Lian}}, \bibinfo
  {author} {\bibfnamefont {D.~A.}\ \bibnamefont {Hales}}, \ and\ \bibinfo
  {author} {\bibfnamefont {P.~B.}\ \bibnamefont {Armentrout}},\ }\href
  {\doibase 10.1021/j100325a001} {\bibfield  {journal} {\bibinfo  {journal} {J.
  Phys. Chem.}\ }\textbf {\bibinfo {volume} {92}},\ \bibinfo {pages} {4009}
  (\bibinfo {year} {1988})}\BibitemShut {NoStop}%
\bibitem [{\citenamefont {Concei\c{c}\~{a}o}\ \emph {et~al.}(1996)\citenamefont
  {Concei\c{c}\~{a}o}, \citenamefont {Loh}, \citenamefont {Lian},\ and\
  \citenamefont {Armentrout}}]{Conceicao96}%
  \BibitemOpen
  \bibfield  {author} {\bibinfo {author} {\bibfnamefont {J.}~\bibnamefont
  {Concei\c{c}\~{a}o}}, \bibinfo {author} {\bibfnamefont {S.~K.}\ \bibnamefont
  {Loh}}, \bibinfo {author} {\bibfnamefont {L.}~\bibnamefont {Lian}}, \ and\
  \bibinfo {author} {\bibfnamefont {P.~B.}\ \bibnamefont {Armentrout}},\ }\href
  {\doibase http://dx.doi.org/10.1063/1.471253} {\bibfield  {journal} {\bibinfo
   {journal} {J. Chem. Phys.}\ }\textbf {\bibinfo {volume} {104}},\ \bibinfo
  {pages} {3976} (\bibinfo {year} {1996})}\BibitemShut {NoStop}%
\bibitem [{\citenamefont {Wigner}\ and\ \citenamefont
  {Witmer}(1928)}]{Wigner28}%
  \BibitemOpen
  \bibfield  {author} {\bibinfo {author} {\bibfnamefont {E.}~\bibnamefont
  {Wigner}}\ and\ \bibinfo {author} {\bibfnamefont {E.~E.}\ \bibnamefont
  {Witmer}},\ }\href {\doibase 10.1007/BF01400247} {\bibfield  {journal}
  {\bibinfo  {journal} {Z. Phys.}\ }\textbf {\bibinfo {volume} {51}},\ \bibinfo
  {pages} {859} (\bibinfo {year} {1928})}\BibitemShut {NoStop}%
\bibitem [{\citenamefont {Crocombette}, \citenamefont {Thole},\ and\
  \citenamefont {Jollet}(1996)}]{Crocombette96a}%
  \BibitemOpen
  \bibfield  {author} {\bibinfo {author} {\bibfnamefont {J.~P.}\ \bibnamefont
  {Crocombette}}, \bibinfo {author} {\bibfnamefont {B.~T.}\ \bibnamefont
  {Thole}}, \ and\ \bibinfo {author} {\bibfnamefont {F.}~\bibnamefont
  {Jollet}},\ }\href {\doibase 10.1088/0953-8984/8/22/013} {\bibfield
  {journal} {\bibinfo  {journal} {J. Phys.: Condens. Matter}\ }\textbf
  {\bibinfo {volume} {8}},\ \bibinfo {pages} {4095} (\bibinfo {year}
  {1996})}\BibitemShut {NoStop}%
\bibitem [{\citenamefont {Piamonteze}, \citenamefont {Miedema},\ and\
  \citenamefont {de~Groot}(2009)}]{Piamonteze09}%
  \BibitemOpen
  \bibfield  {author} {\bibinfo {author} {\bibfnamefont {C.}~\bibnamefont
  {Piamonteze}}, \bibinfo {author} {\bibfnamefont {P.}~\bibnamefont {Miedema}},
  \ and\ \bibinfo {author} {\bibfnamefont {F.~M.~F.}\ \bibnamefont
  {de~Groot}},\ }\href {\doibase 10.1103/PhysRevB.80.184410} {\bibfield
  {journal} {\bibinfo  {journal} {Phys. Rev. B}\ }\textbf {\bibinfo {volume}
  {80}},\ \bibinfo {pages} {184410} (\bibinfo {year} {2009})}\BibitemShut
  {NoStop}%
\bibitem [{\citenamefont {Komelj}\ \emph {et~al.}(2002)\citenamefont {Komelj},
  \citenamefont {Ederer}, \citenamefont {Davenport},\ and\ \citenamefont
  {F\"ahnle}}]{Komelj02}%
  \BibitemOpen
  \bibfield  {author} {\bibinfo {author} {\bibfnamefont {M.}~\bibnamefont
  {Komelj}}, \bibinfo {author} {\bibfnamefont {C.}~\bibnamefont {Ederer}},
  \bibinfo {author} {\bibfnamefont {J.~W.}\ \bibnamefont {Davenport}}, \ and\
  \bibinfo {author} {\bibfnamefont {M.}~\bibnamefont {F\"ahnle}},\ }\href
  {\doibase 10.1103/PhysRevB.66.140407} {\bibfield  {journal} {\bibinfo
  {journal} {Phys. Rev. B}\ }\textbf {\bibinfo {volume} {66}},\ \bibinfo
  {pages} {140407} (\bibinfo {year} {2002})}\BibitemShut {NoStop}%
\bibitem [{\citenamefont {Ederer}\ \emph {et~al.}(2003)\citenamefont {Ederer},
  \citenamefont {Komelj}, \citenamefont {Davenport},\ and\ \citenamefont
  {F\"ahnle}}]{Ederer03}%
  \BibitemOpen
  \bibfield  {author} {\bibinfo {author} {\bibfnamefont {C.}~\bibnamefont
  {Ederer}}, \bibinfo {author} {\bibfnamefont {M.}~\bibnamefont {Komelj}},
  \bibinfo {author} {\bibfnamefont {J.~W.}\ \bibnamefont {Davenport}}, \ and\
  \bibinfo {author} {\bibfnamefont {M.}~\bibnamefont {F\"ahnle}},\ }\href@noop
  {} {\bibfield  {journal} {\bibinfo  {journal} {J. Electron Spectrosc. Relat.
  Phenom.}\ }\textbf {\bibinfo {volume} {130}},\ \bibinfo {pages} {97}
  (\bibinfo {year} {2003})}\BibitemShut {NoStop}%
\bibitem [{\citenamefont {Ederer}, \citenamefont {Komelj},\ and\ \citenamefont
  {F\"ahnle}(2003)}]{Ederer03a}%
  \BibitemOpen
  \bibfield  {author} {\bibinfo {author} {\bibfnamefont {C.}~\bibnamefont
  {Ederer}}, \bibinfo {author} {\bibfnamefont {M.}~\bibnamefont {Komelj}}, \
  and\ \bibinfo {author} {\bibfnamefont {M.}~\bibnamefont {F\"ahnle}},\ }\href
  {\doibase 10.1103/PhysRevB.68.052402} {\bibfield  {journal} {\bibinfo
  {journal} {Phys. Rev. B}\ }\textbf {\bibinfo {volume} {68}},\ \bibinfo
  {pages} {052402} (\bibinfo {year} {2003})}\BibitemShut {NoStop}%
\bibitem [{\citenamefont {Min\'ar}\ \emph {et~al.}(2006)\citenamefont
  {Min\'ar}, \citenamefont {Bornemann}, \citenamefont {\v{S}ipr}, \citenamefont
  {Polesya},\ and\ \citenamefont {Ebert}}]{Minar06}%
  \BibitemOpen
  \bibfield  {author} {\bibinfo {author} {\bibfnamefont {J.}~\bibnamefont
  {Min\'ar}}, \bibinfo {author} {\bibfnamefont {S.}~\bibnamefont {Bornemann}},
  \bibinfo {author} {\bibfnamefont {O.}~\bibnamefont {\v{S}ipr}}, \bibinfo
  {author} {\bibfnamefont {S.}~\bibnamefont {Polesya}}, \ and\ \bibinfo
  {author} {\bibfnamefont {H.}~\bibnamefont {Ebert}},\ }\href {\doibase
  10.1007/s00339-005-3359-1} {\bibfield  {journal} {\bibinfo  {journal} {Appl.
  Phys. A: Mater. Sci. Process.}\ }\textbf {\bibinfo {volume} {82}},\ \bibinfo
  {pages} {139} (\bibinfo {year} {2006})}\BibitemShut {NoStop}%
\bibitem [{\citenamefont {\v{S}ipr}, \citenamefont {Mi\'nar},\ and\
  \citenamefont {Ebert}(2009)}]{Sipr09a}%
  \BibitemOpen
  \bibfield  {author} {\bibinfo {author} {\bibfnamefont {O.}~\bibnamefont
  {\v{S}ipr}}, \bibinfo {author} {\bibfnamefont {J.}~\bibnamefont {Mi\'nar}}, \
  and\ \bibinfo {author} {\bibfnamefont {H.}~\bibnamefont {Ebert}},\ }\href
  {\doibase 10.1209/0295-5075/87/67007} {\bibfield  {journal} {\bibinfo
  {journal} {Europhys. Lett.}\ }\textbf {\bibinfo {volume} {87}},\ \bibinfo
  {pages} {67007} (\bibinfo {year} {2009})}\BibitemShut {NoStop}%
\bibitem [{\citenamefont {Berdyugina}\ and\ \citenamefont
  {Solanki}(2002)}]{Berdyugina02}%
  \BibitemOpen
  \bibfield  {author} {\bibinfo {author} {\bibfnamefont {S.~V.}\ \bibnamefont
  {Berdyugina}}\ and\ \bibinfo {author} {\bibfnamefont {S.~K.}\ \bibnamefont
  {Solanki}},\ }\href {\doibase 10.1051/0004-6361:20020130} {\bibfield
  {journal} {\bibinfo  {journal} {Astron. Astrophys.}\ }\textbf {\bibinfo
  {volume} {385}},\ \bibinfo {pages} {701} (\bibinfo {year}
  {2002})}\BibitemShut {NoStop}%
\bibitem [{\citenamefont {{Asensio Ramos}}\ and\ \citenamefont {{Trujillo
  Bueno}}(2006)}]{AsensioRamos06}%
  \BibitemOpen
  \bibfield  {author} {\bibinfo {author} {\bibfnamefont {A.}~\bibnamefont
  {{Asensio Ramos}}}\ and\ \bibinfo {author} {\bibfnamefont {J.}~\bibnamefont
  {{Trujillo Bueno}}},\ }\href {\doibase 10.1086/497892} {\bibfield  {journal}
  {\bibinfo  {journal} {Astrophys. J.}\ }\textbf {\bibinfo {volume} {636}},\
  \bibinfo {pages} {548} (\bibinfo {year} {2006})}\BibitemShut {NoStop}%
\bibitem [{\citenamefont {Asher}\ \emph {et~al.}(1994)\citenamefont {Asher},
  \citenamefont {Bellert}, \citenamefont {Buthelezi},\ and\ \citenamefont
  {Brucat}}]{Asher94a}%
  \BibitemOpen
  \bibfield  {author} {\bibinfo {author} {\bibfnamefont {R.}~\bibnamefont
  {Asher}}, \bibinfo {author} {\bibfnamefont {D.}~\bibnamefont {Bellert}},
  \bibinfo {author} {\bibfnamefont {T.}~\bibnamefont {Buthelezi}}, \ and\
  \bibinfo {author} {\bibfnamefont {P.}~\bibnamefont {Brucat}},\ }\href
  {\doibase http://dx.doi.org/10.1016/0009-2614(94)00573-7} {\bibfield
  {journal} {\bibinfo  {journal} {Chem. Phys. Lett.}\ }\textbf {\bibinfo
  {volume} {224}},\ \bibinfo {pages} {525} (\bibinfo {year}
  {1994})}\BibitemShut {NoStop}%
\bibitem [{\citenamefont {Yang}\ and\ \citenamefont {Hackett}(2000)}]{Yang00b}%
  \BibitemOpen
  \bibfield  {author} {\bibinfo {author} {\bibfnamefont {D.-S.}\ \bibnamefont
  {Yang}}\ and\ \bibinfo {author} {\bibfnamefont {P.~A.}\ \bibnamefont
  {Hackett}},\ }\href {\doibase
  http://dx.doi.org/10.1016/S0368-2048(99)00073-0} {\bibfield  {journal}
  {\bibinfo  {journal} {J. Electron Spectrosc. Relat. Phenom.}\ }\textbf
  {\bibinfo {volume} {106}},\ \bibinfo {pages} {153} (\bibinfo {year}
  {2000})}\BibitemShut {NoStop}%
\bibitem [{\citenamefont {Lefebvre-Brion}\ and\ \citenamefont
  {Field}(1986)}]{LefebvreBrion86}%
  \BibitemOpen
  \bibfield  {author} {\bibinfo {author} {\bibfnamefont {H.}~\bibnamefont
  {Lefebvre-Brion}}\ and\ \bibinfo {author} {\bibfnamefont {R.~W.}\
  \bibnamefont {Field}},\ }\href@noop {} {\emph {\bibinfo {title}
  {Perturbations in the Spectra of Diatomic Molecules}}}\ (\bibinfo
  {publisher} {Academic Press},\ \bibinfo {address} {Orlando},\ \bibinfo {year}
  {1986})\BibitemShut {NoStop}%
\bibitem [{\citenamefont {Egashira}\ \emph {et~al.}(2015)\citenamefont
  {Egashira}, \citenamefont {Yamada}, \citenamefont {Kita},\ and\ \citenamefont
  {Tachikawa}}]{Egashira15}%
  \BibitemOpen
  \bibfield  {author} {\bibinfo {author} {\bibfnamefont {K.}~\bibnamefont
  {Egashira}}, \bibinfo {author} {\bibfnamefont {Y.}~\bibnamefont {Yamada}},
  \bibinfo {author} {\bibfnamefont {Y.}~\bibnamefont {Kita}}, \ and\ \bibinfo
  {author} {\bibfnamefont {M.}~\bibnamefont {Tachikawa}},\ }\href {\doibase
  10.1063/1.4907197} {\bibfield  {journal} {\bibinfo  {journal} {J. Chem.
  Phys.}\ }\textbf {\bibinfo {volume} {142}},\ \bibinfo {eid} {054309}
  (\bibinfo {year} {2015})}\BibitemShut {NoStop}%
\end{thebibliography}
%

\end{document}